\documentclass[aps,reprint,amsmath,amssymb,longbibliography,nofootinbib]{revtex4-2}
\usepackage[T1]{fontenc}
\usepackage{mathrsfs,booktabs,array,microtype,hyperref}
\hypersetup{hidelinks,pdftitle={Curved-spacetime pNRQCD}}
\allowdisplaybreaks[2]
\newcommand{\dd}{\,\mathrm d}
\newcommand{\ii}{\mathrm i}
\newcommand{\as}{\alpha_s}
\newcommand{\CF}{C_F}
\newcommand{\CA}{C_A}
\newcommand{\Tf}{T_F}
\newcommand{\T}{\mathcal T}
\newcommand{\U}{\mathcal U}
\newcommand{\Herm}{\operatorname{Herm}}
\newcommand{\tr}{\operatorname{tr}}

\newcommand{\EB}{E_B}

\newcommand{\FP}{\operatorname{FP}}
\newcommand{\one}{\mathbf 1}
\begin{document}
\title{Curved-spacetime pNRQCD}
\author{Michael Wilson}
\email{mkwilson3@ualr.edu}
\affiliation{Department of Physics, University of Arkansas at Little Rock, Little Rock, Arkansas, USA}
\begin{abstract}
We formulate potential nonrelativistic QCD for a heavy quark--antiquark pair on a prescribed classical spacetime. In the static limit the curvature dependence of the interaction is determined by the matching of the singlet and projected-octet potentials, which we calculate through one loop in a static zero-shift geometry. For a mobile pair the gravitational rest-mass interaction also enters the chromoelectric dipole self-energy. Although common to the singlet and octet, this interaction modifies the ultrasoft running at the same order as the geometric correction to the chromoelectric correlator. We obtain the corresponding relative two-dipole counterterms and formulate the real-time kernel without an ultrasoft adiabatic approximation. On ultrastatic slices we calculate singlet level differences through $O(\alpha_s R/M)$ in the spin and center-rest projections. A weighted combination of magnetic-sublevel splittings isolates the radiative curvature contribution. Its dependence on the number of massless light flavors at fixed matched mass, coupling and geometry is determined by the QCD beta function.
\end{abstract}
\maketitle
\section{Introduction}
In potential nonrelativistic QCD (pNRQCD), the separation of the relative-momentum and binding-energy scales permits the static color interaction to be treated as a Wilson coefficient. Its infrared sensitivity is canceled by the ultraviolet sensitivity of the chromoelectric dipole self-energy \cite{bpsv,review,pineda}. On a classical gravitational background the propagation of both the heavy constituents and the ultrasoft gluons depends on the geometry.

In a Ricci-flat background the electric Weyl tensor introduces a \(t^{-2}\) term in the short-time chromoelectric correlator, giving rise to a curvature-dependent ultrasoft logarithm at order \(\alpha_s^2\). The static singlet running and its finite optical matching obtained in \cite{wilson} are recovered here as a special case. For a mobile pair the gravitational rest-mass interaction enters the propagation of both the singlet and octet. Although it leaves the static color gap unchanged, its noncommutativity with the relative kinetic energy gives an additional contribution to the cubic dipole self-energy. Both geometric contributions enter the ultrasoft running at the same order. We calculate the mobile anomalous dimension together with the static matching for Ricci curvature, acceleration and the projected octet.

The geometric dependence of the heavy-pair Hamiltonian also enters the bound-state spectrum. On ultrastatic slices the spatial curvature modifies the kinetic energy and the color interaction in the absence of an electric mass tide. This permits a calculation of the radiative curvature dependence of the spectrum without the general finite matching of the mobile Hamiltonian.

QCD on classical gravitational backgrounds has been studied in lattice formulations \cite{yamamoto} and in effective descriptions of non-Abelian plasma screening \cite{alonsokaiser}. The present target is different: the soft matching and ultrasoft renormalization of a nonrelativistic heavy pair, rather than a plasma screening action, a holographic potential \cite{holographic}, or a phenomenological confinement model \cite{conical,gaete}. Microscopic stress-tensor couplings of gluons have been studied in connection with the trace anomaly \cite{armillis}. Gravitational two-body Hamiltonians, Fermi-frame reductions and radiative bound-state responses have Abelian and one-body precedents \cite{caicedo,percheneuser,perche,ji_et_al}. The curved Coulomb problem itself is established \cite{redkov}. Here the radiative corrections are obtained from the non-Abelian matching and the resolved octet spectrum.

The cubic dipole self-energy and its flat-space operator reduction are known \cite{pineda2011}; we calculate their geometric variation together with the change in the field-strength correlator. The latter uses the curved vector parametrix \cite{frob,vass}. Thermal and open-system pNRQCD provide precedents for spectral moments and closed-time-path propagation \cite{thermal,oqs,oqs2026}. The Wilson first-recoil construction \cite{firstrecoil,secondrecoil} is used to delimit the stationary accuracy, and the conformal relation of the cusp to the cylinder static energy \cite{cusp,cuspreview} provides a special-geometry check. The ultraviolet singularity considered here is local and state independent for the stipulated Hadamard states, whereas the finite real-time response depends on the preparation and light-field state.

The construction and its power counting are given in Sec.~\ref{sec:action}. Section~\ref{sec:soft} matches the local static potentials, including the infrared overlap. In Sec.~\ref{sec:us} we derive the electric correlator and the static and mobile ultrasoft counterterms, and formulate their nonstationary extension. Section~\ref{sec:spectra} combines the soft and ultrasoft results in two stationary spectroscopic observables. The separate accuracy of the pinned, mobile and spectral sectors is fixed in Sec.~\ref{sec:geom}; finite matching for an arbitrary mobile observable is not inferred from local running. The appendices contain the basis reduction, an independent gauge calculation, and auxiliary spectral transforms.
\section{Effective theory}\label{sec:action}
\subsection{Geometry and scale separation}\label{sec:geom}
We set $\hbar=c=1$, use signature $(-,+,+,+)$ and $[\nabla_\mu,\nabla_\nu]V^\rho=R^\rho{}_{\sigma\mu\nu}V^\sigma$. Let $X(\tau)$ have unit tangent $u$ and acceleration $a=\nabla_u u$. In a nonrotating Fermi--Walker tetrad, write $r^\mu\in u^\perp$, $r=(r^\mu r_\mu)^{1/2}$ and $n^\mu=r^\mu/r$. The endpoints are
\begin{equation}
 \begin{gathered}x_\pm=\exp_X(\pm r/2),\quad \gamma_\sigma=\exp_X(\sigma r),\\ -1/2\leq\sigma\leq1/2.\end{gathered}
\end{equation}
We work in a normal tube in which the spatial spokes are unique. The common time is midpoint proper time, not the proper time of either endpoint. The acceleration and electric tidal tensor are $a_i$ and $\T_{ij}=R_{i0j0}$; to the required Fermi order,
\begin{align}
 g_{00}&=-(1+a_i y^i)^2-\T_{ij}y^iy^j+\cdots,\\
 g_{0i}&=-\tfrac23 R_{0jik}y^jy^k+\cdots,\nonumber\\
 g_{ij}&=\delta_{ij}-\tfrac13 R_{ikjl}y^ky^l+\cdots .
\end{align}
In particular $N_\pm=1\pm a\cdot r/2+\T_{ij}r^ir^j/8+\cdots$. For two equal kinetic masses $M$, the common rest-mass tide after removing the rest monopole is
\begin{equation}\label{eq:tide}
 F=\frac M4\T_{ij}r^ir^j.
\end{equation}
It is not part of the isolated-line-subtracted interaction potential.

The hard and soft matching hierarchy is
\begin{equation}\label{eq:hier}
 \begin{gathered}M\gg Q=r^{-1}\gg\{k_{\rm us},\Lambda_{\rm QCD},\Omega_{\rm geom},|a|,\\\sqrt{\|R_{\mu\nu\rho\sigma}\|}\},\qquad \as(Q)\ll1.\end{gathered}
\end{equation}
Spatial gradients on the pair and temporal gradients at the soft scale are expanded. Near zeros of curvature the derivative hierarchy is stated in terms of absolute jets, not their ratio. In Coulombic applications $a_0=2/(M\CF\as)$, $v=(Ma_0)^{-1}$ and $\EB=Mv^2$. For a physical perturbative ultrasoft spectrum we also take $\Lambda_{\rm QCD}\ll\EB$. For the stationary short-time reference used to obtain physical adiabatic energies we also require $\sqrt{\|R\|},\Omega_{\rm geom}\ll\EB$. This stronger condition is not needed for the real-time functional. We retain a single power of curvature or two of acceleration in local static terms, and the leading two electric dipoles in ultrasoft loops. The static coefficients apply to local static zero-shift completions; the local electric singularity applies to general smooth nonrotating lines. In the mobile sector we retain the leading relative kinetic term, the common mass tide and the leading two-dipole geometric pole. We do not match every finite recoil or higher-multipole coefficient. For bound-state perturbation theory $\|\T\|/E_B^2\ll1$ is needed independently of $\|\T\|a_0^2\ll1$; $F/E_B$ at $r\sim a_0$ is of order $\|\T\|/(4E_B^2)$. Acceleration must satisfy $|a|a_0\ll1$ for local soft matching, and $|a|/E_B\ll1$ if an accelerated short-time expansion is used in a stationary adiabatic energy calculation. The external support and field state are part of the pinned protocol. In a mobile matching patch, the common energy change must also be small relative to the eliminated soft energy: a sufficient condition is $M\|\T\|r^2\ll Q$ (and $M|a|r\ll Q$ for endpoint lapse gradients). In the Coulombic counting for the leading geometric pole below take relative momenta of order $Mv$ and resolved energies of order $E_B$, including temporal geometry variations at most of that order. The pole is a linear-geometry, leading-velocity result; the time-ordered kernel itself does not require a slow variation relative to $E_B$. For the formal stationary perturbative coefficients take the curvature expansion in the perturbative vacuum; a physical perturbative realization can use $\Lambda_{\rm QCD}\ll q\ll E_B$ (with $q$ a curvature momentum) and a weak coupling at all retained dynamical scales. The limit $q\to0$ at fixed nonzero confinement scale need not be described by this expansion.
\paragraph{Order of the matching.} The perturbative accuracy is specified separately for the observables and local kernels. The diagonal pinned coefficients retain $M^0$, geometric weight two, and terms through $\as^2$. The mobile construction retains the tree kinetic lift and the common tide, and its local loop matching is the leading relative two-E1 pole at order $g^2$. The finite real-time functional at that order uses the supplied light-field state and the retained heavy propagators; it is not a matching of independent finite transverse or gravitational source amplitudes. In Sec.~\ref{sec:spectra} we set $\T=0$ and calculate differences through $O(\as R/M)$ with $\as\sim v$; $R$ in this order estimate denotes a curvature scale. The radiative separated first-recoil coefficient $O(\as^2R/M)$, and radiative corrections to the leading E1 vertices in the geometric loop, are then beyond the retained order. Independent hard Ricci scalars can instead be of the retained order, but cancel in these homogeneous level differences. This does not assert uniform finite matching of the general mobile Hamiltonian to that accuracy when $\T\ne0$.
\subsection{Curved-spacetime NRQCD}
We use $D=\nabla+\ii g A$ and $F_{\mu\nu}=[D_\mu,D_\nu]_{\rm color}/(\ii g)$. In a static zero-shift metric $\dd s^2=-N^2\dd t^2+h_{ij}\dd x^i\dd x^j$, let $\Pi=-\ii D_i$ on an endpoint. A tree-level static-normal Foldy--Wouthuysen representative is
\begin{equation}\label{eq:hard}
 \begin{aligned}H_{\rm end}={}&MN+g A_t+\frac1{2M}\bigl[\Pi_iN\Pi^i+Ng\boldsymbol\sigma\cdot\boldsymbol B\\
 &+\tfrac12\boldsymbol\sigma\cdot(\nabla N\times\boldsymbol\Pi)+\tfrac14(N{}^{(3)}R-\Delta_hN)\bigr].\end{aligned}
\end{equation}
Indeed the odd Dirac operator is $\{N,\alpha^i\Pi_i\}/2$. Its first positive-energy even term is $\{\alpha\cdot\Pi,\{N,\alpha\cdot\Pi\}\}/(8M)$, which gives Eq.~\eqref{eq:hard} with the invariant adjoints. In the conjugate representation use the conjugate generators. Radiative coefficients of the dimension-five operators and general time-dependent hard matching are distinct from this tree reduction.

\subsection{Color and geometric variables}
The light action is the generally covariant Yang--Mills and massless Dirac action. Pure metric counterterms of engineering dimension at most four may be required for the vacuum functional; they cancel in vacuum-normalized pair observables at a fixed background. We transport the endpoint color indices to the midpoint along the spokes. The pair matrix $\mathcal P=S\one/\sqrt{N_c}+O^a T^a/\sqrt{\Tf}$ transforms by conjugation there. We use $\Tf=1/2$, $\CF=(N_c^2-1)/(2N_c)$, $\CA=N_c$, $g^2=4\pi\as$, $X_s=-\CF$, $X_o=\CA/2-\CF$. Let $P_s,P_o$ be the projectors on these two color components.

With proper time at the midpoint as the time variable, the action is
\begin{equation}\label{eq:action}
 \begin{aligned}
 I={}&I_{\rm light}+\int\dd\tau\,\dd\mu_{\rm pair}\,\tr\{\mathcal P^\dagger(\ii D_\tau-\widehat H_{\rm diag})\mathcal P\}\\
 &-\int\dd\tau\,\dd\mu_{\rm pair}\,\mathcal H_{\rm mult} .
 \end{aligned}
\end{equation}
Here $D_\tau$ includes the horizontal frame and gauge transports, the singlet has no adjoint connection, and the color trace in the diagonal term includes the indicated projectors. The pair measure and kinetic operator are given below for a stationary mobile pair. For prescribed sources the center and separation are fixed, and no integration over their dynamical variables is needed. The time-dependent mobile action is understood with its induced slice measure, or equivalently in half-densities. In the pinned limit the diagonal operator is
\begin{equation}
 \widehat H_{\rm diag}=\sum_{a=s,o}P_a\bigl[V_a^{\rm flat}(r;\mu,\nu)+V_a^{\rm geom}(X,r;\mu,\nu)\bigr],
\end{equation}
where the isolated rest phases have been removed. Define $L_\mu=\ln(\mu r)+\gamma_E$, $\beta_0=(11\CA-4\Tf n_f)/3$, and $a_1=(31\CA-20\Tf n_f)/9$. At one loop $V_a^{\rm flat}=X_a\as[1+\as(a_1+2\beta_0 L_\mu)/(4\pi)]/r$ and $V_a^{\rm geom}$ is given in Sec.~\ref{sec:soft}. The action is Hermitian before integrating out the ultrasoft fields. It does not contain a complex static potential intended to replace their real-time propagation.

For pinned sources, let $E_i=F_{i\mu}u^\mu$ and $\mathscr S=S\one/\sqrt{N_c}$, $\mathscr O=O^aT^a/\sqrt{\Tf}$. The leading electric multipoles in the Hamiltonian convention are
\begin{equation}
 \begin{aligned}\mathcal H_{E1}={}&g\tr\bigl[V_A(\mathscr O^\dagger r^iE_i\mathscr S+\mathrm{h.c.})\\
 &+\tfrac12 V_B\mathscr O^\dagger\{r^iE_i,\mathscr O\}\bigr],\\
 &V_A=V_B=1\quad\text{at tree level}.\end{aligned}
\end{equation}
The sign of the off-diagonal vertex can equivalently be fixed by rephasing the octet; the squared dipole color factors used below are unaffected. To classify single local pinned insertions, count $D_{\rm us},a,\Delta$ as weight one, $gF,R_{\mu\nu\rho\sigma}$ as weight two and light bilinears as weight three. At weight two the six scalars are
\begin{equation}\label{eq:slots}
 Z=(R,R_{uu},R_{nn},C_{nunu},a^2,a_n^2).
\end{equation}
At weight three the remaining one-field operators for prescribed sources can be grouped as follows. For adjoint Hermitian $Y$, define $\mathcal F(Y)=\tr(\mathscr O^\dagger[Y,\mathscr O])$ and $\mathcal P_-(Y)=\ii\tr(\mathscr S^\dagger Y\mathscr O-\mathscr O^\dagger Y\mathscr S)$. Up to the symmetry and local EOM reductions of the pinned basis, the two groups of terms are $r^2\mathcal F(Y)$ and $r^2\mathcal P_-(Y)$ with arguments
\begin{align}
 \mathcal F:&\quad gD_iE_i,\quad gn^in^jD_iE_j,\quad ga^iE_i,\quad ga_nE_n,\quad J_u,\\
 \mathcal P_-:&\quad gn^iD^j F_{ij},\quad gn^ia^jF_{ij},\quad J_n,
\end{align}
where $J^\mu=g\sum_f(\bar q_f\gamma^\mu T^a q_f)T^a$. Derivatives of $F$ are projected after differentiation. We do not use the light equations of motion to identify these source insertions. Including two flat diagonal terms, twelve geometric diagonal terms and the two dipoles, the basis contains 24 operators (22 in pure Yang--Mills). The classification assumes $\theta_{\rm QCD}=0$, $N_c\geq3$, and chiral flavor symmetry (or a chirality-preserving perturbative calculation for one light flavor). It is a classification of single pinned local insertions in a canonical off-shell light-source convention, not a general mobile basis. Appendix~\ref{app:basis} gives the reduction. The finite radiative coefficients of the weight-three one-field operators are not fixed here and are not needed for the retained leading two-dipole loop.
\subsection{Kinetic lift and stationary recoil}
For the spatial midpoint map $\varphi:(X,r)\mapsto(x_+,x_-)$, pull back the endpoint volume and inverse metric tensors:
\begin{align}
 \dd\mu_{\rm pair}&=\varphi^*(\dd\operatorname{vol}_h(x_+)\dd\operatorname{vol}_h(x_-)),\\
 \mathbb A&=(\varphi_*)^{-1}(N_+h_+^{-1}\oplus N_-h_-^{-1})(\varphi_*)^{-T},\\
 H_{\rm kin}&=-\frac1{2M}\operatorname{div}_{\dd\mu_{\rm pair}}^\varphi\mathbb A\nabla^\varphi .\label{eq:kin}
\end{align}
Spoke differentiation adds to the midpoint derivative on each endpoint the transported field-strength integral $g\int_0^{\pm1/2}\dd\sigma\,U(X,\gamma_\sigma)F_{\mu\nu}(\gamma_\sigma)U(\gamma_\sigma,X)\partial_\sigma\gamma^\mu\partial_A\gamma^\nu$. It acts from the left or right according to the endpoint color. This construction fixes the minimal recoil vertices and the spin transport.

Writing $\mathscr R_r v=R^h(v,r)r$, the endpoint Jacobi maps are $A v\pm B w$, $A=1-\mathscr R_r/8$, $B=(1-\mathscr R_r/24)/2$. The Jacobian is $1-\operatorname{Ric}_h(r,r)/6$. If $P,p$ denote center and relative momenta, the principal symbol per unit midpoint time is
\begin{equation}
 \frac{1+\T_{rr}/8}{4M}(P^2+4p^2)+\frac{a\cdot r}{2M}P\cdot p+\frac{P\cdot\mathscr R_r P}{16M}+\frac{p\cdot\mathscr R_r p}{12M}.
\end{equation}
The subprincipal terms follow from the measure and divergence form in Eq.~\eqref{eq:kin}; the principal symbol alone does not determine them. In the normalized Wilson representation at one hard recoil insertion the principal kinetic coefficient is unchanged by the static soft projection: for the isometric embedding $I$, the derivative $\nabla(I\psi)=I\nabla_{\rm low}\psi+(1-II^\dagger)(\nabla I)\psi$ is an orthogonal decomposition. The second term supplies conservative multiplication operators. The stationary field-free Hamiltonian in Killing time consists of the endpoint rest, kinetic and spin/scalar terms, $N(X)V_a^{(0)}$, and
\begin{equation}\label{eq:recoilslots}
 \frac{N(X)}{M}\sum_{a=s,o}P_a\left[V_a^{(1)}(r)+\sum_{k=1}^6 d_{ak}(r)Z_k\right].
\end{equation}
Equation~\eqref{eq:recoilslots} contains fourteen first-recoil scalar operators. Their finite coefficients are not all matched for arbitrary $Z_k$. The undetermined functions $d_{ak}$ cannot in general be set to zero. The ultrastatic level differences below do not require these functions through their stated orders; the power-counting argument is given in Sec.~\ref{app:spectral}.
\section{Matching the potentials}\label{sec:soft}
\subsection{Definition of the matching and tree potential}
We match in an auxiliary static completion with the physical midpoint lapse one, prescribed endpoints, common isolated-line subtraction and compatible gapped infrared reference. The optical spatial metric is $\widetilde h=N^{-2}h$. At the midpoint write $S_{ij}=\widetilde R_{ij}$, $S=\tr S_{ij}$, $P=S_{nn}-S/3=2C_{nunu}$ and $K=S/6=R/6+R_{uu}-a^2$. Continue $d=3-2\epsilon$ with the traceless projection $P$ held fixed and $S^{\rm tr}_{ij}=(d-1)K\delta_{ij}$. Write $b=(d-1)/2$, $c_b=\Gamma(b)/(4\pi^{b+1})$, and $\ell$ for optical endpoint distance. In the optical parallel frame let $z$ be the separation vector of length $\ell$ and $S_{zz}=S_{ij}z^iz^j$. To quadratic geometric order,
\begin{equation}
 \ell=r(1+r^2 B_\ell),\quad B_\ell=(3a_n^2-a^2-\T_{nn})/24.
\end{equation}
The bending of the optical segment is included in this formula. The scalar fractional Green functions are, through linear optical curvature,
\begin{equation}\label{eq:frac}
 \begin{gathered}G_j=k_j\ell^{2j-d}\left[1+\frac{S_{zz}}{12}+\frac{S\ell^2}{24(d/2-j-1)}\right],\\
 k_j=\frac{\Gamma(d/2-j)}{4^j\pi^{d/2}\Gamma(j)}.\end{gathered}
\end{equation}
Thus the tree coefficient of $X_a\as r$ is $(S_{nn}-S)/12-B_\ell$. Decompose $\T_{ij}=C_{i0j0}+(R_{uu}\delta_{ij}-R_{ij})/2+R\delta_{ij}/6$ to express it in Eq.~\eqref{eq:slots}.

For the one-loop terms, the local temporal and ghost Green functions and spatial Hodge Green function are, with $Z_t=t^2+\ell^2$, $U=1+S_{zz}/12$,
\begin{align}
 H&=Uc_b Z_t^{-b}+\frac{c_b S}{24(b-1)}Z_t^{1-b},\\
 D_{ij'}&=g_{ij'}H-\frac{c_b S_{ij'}}{4(b-1)}Z_t^{1-b}.
\end{align}
\subsection{One-loop polarization and connected Wilson web}\label{app:loops}
The integrated scalar-covariantized polarization is proportional to $L^b$, where $L=-\widetilde\nabla^2$. The dimensionally continued tree normalization is used throughout.
The trace-free attachment factor $\mathcal A(b)=b(b-1)/[(2-b)(3-b)]$ follows from applying $-\partial^2(S_{zz}\ell^p)=-p(p+d+2)S_{zz}\ell^{p-2}$ twice. For the trace continuation define \(q_j\) so that Eq.~\eqref{eq:frac} has relative curvature \(K\ell^2 q_j\). If an integrated polarization has relative coefficient \(K\ell^2 R_p\), its attached and tree-divided correction is
\begin{align}
 q_j&=\frac{d-1}{12}+\frac{d(d-1)}{24(d/2-j-1)},\nonumber\\
 \rho_p^{\rm tr}
 &=q_{2-b}-q_1+(R_p-q_{-b})\frac{k_{3-b}k_{-b}}{k_{2-b}k_{1-b}} .\label{eq:attachTr}
\end{align}
This attaches \(L^{b-1}\) to the residual trace term. In both channels the overall normalization is the dimensionally continued tree through the retained geometric order. The ordinary coupling subtraction removes its \(\beta_0/\epsilon\) multiple before expanding any evanescent tree factor.

\subsubsection{Gluon, ghost and fermion polarization}
Split an external static temporal field \(B^a\) from the quantum fields, which are denoted by \(A\) in this subsection. The coefficient of one \(B^a\) in the cubic action, after integrating its spatial derivative, is proportional to
\[
 J_0^a=f^{abc}\left[-\dot A_i^b A_i^c
                    +2(\widetilde\nabla_i A_0^b)A_i^c
                    +A_0^b\widetilde\nabla_i A_i^c\right].
\]
An overall sign convention for \(J_0^a\) drops out. The three terms are \(J_T,J_M,J_D\). Their static contraction and the ghost contribute the following integrand, after dividing out \(\CA\):
\begin{equation}\label{eq:current}
\begin{aligned}
 {\cal P}_g=\int_{-\infty}^\infty\dd t\,\bigl[&2\dot D_{ij'}\dot D_{ij'}+4D_{ij'}H_{;ij'}\\
 &-2|\widetilde\nabla_xH|^2-2|\widetilde\nabla_yH|^2-2\dot H^2\bigr].
\end{aligned}
\end{equation}
The \(TT\) contraction gives one \(\dot D^2\) directly and another by time integration by parts. \(MM\) gives the mixed Hessian term. \(MD,DM\) each use \(\widetilde\nabla_x^iD_{ij'}=-\widetilde\nabla_{j'}H\). \(DD\) gives \(H(\partial_t^2 H+\delta)\), and the Grassmann ghost supplies the second \(-\dot H^2\) after the time integral. The contact \(H\delta\) and the \(B^2A_i^2\) tadpole are local and scaleless for the homogeneous all-hard parametrix. Their distinct regulated contributions are kept in Sec.~\ref{app:overlap}. This derivation uses ordinary Feynman gauge; using background gauge without changing the vertices in Eq.~\eqref{eq:current} would change the result.

Put \(j_d=\sqrt{\pi}\Gamma(d-1/2)/\Gamma(d)\) and \(A_b=4b^2(3+1/d)\). The flat kernel after the time integral is \(-c_b^2j_d A_b\ell^{1-2d}\). For trace-free curvature, the mixed Hessian and gradient norm, after pulling the indices into parallel frames, are
\begin{align*}
 g^{ij'}H_{;ij'}&=U(-2dF'-4\ell^2F'')-S_{zz}F',\\
 |\widetilde\nabla_x H|^2&=4\ell^2 U^2(F')^2+(2/3)S_{zz}F F'.
\end{align*}
Here $Z=Z_t$ and $F=c_b Z^{-b}$; primes are $Z$-derivatives. The first curvature term in the Hessian comes from the Jacobi correction to \(Z_{;ij'}\) and the two differentiated Van Vleck factors. The \(D\)-endomorphism in \(4D H_{;ij'}\) gives \(4c_b S_{zz}Z^{1-b}F''/(b-1)\). Thus the traceless curvature numerator beyond \(U^2\) is \(c_b^2 S_{zz}Z^{-d}\,B_b\), where
\[
 B_b=20b/3+4b(b+1)/(b-1).
\]
In the trace continuation use \(U=1+A K\ell^2\), \(A=(d-1)/12\), and write \(H=F[U+B KZ]\), \(D=g_{ij'}F[U+C KZ]\), where
\[
 B=\frac{d(d-1)}{24(b-1)},\qquad C=B-\frac{d-1}{4(b-1)} .
\]
The mixed Hessian trace is \(-H_{\ell\ell}-(d-1)(1/\ell+K\ell/6)H_\ell\). This follows from the transverse mixed Jacobi Hessian \(-1/S_K(\ell)\), where \(S_K''=-K S_K\), \(S_K(0)=0\), \(S_K'(0)=1\). Substitution in Eq.~\eqref{eq:current} gives the trace polarization. All terms reduce by
\begin{equation}\label{eq:moments}
 \int_{\mathbb R}\dd t\,\frac{t^{2m}}{(t^2+\ell^2)^p}
 =\frac{\Gamma(m+1/2)\Gamma(p-m-1/2)}{\Gamma(p)}
    \ell^{2m+1-2p}.
\end{equation}
The resulting pre-attachment correction is \(K\ell^2 R_g\), with
\begin{equation}\label{eq:tracebubble}
 R_g=\frac{24b^3-11b^2-3b+5}{3(3b+2)(4b-1)} .
\end{equation}

For the light fermions the squared Dirac operator is \(-\nabla^2+S/4\), giving heat endomorphism \(-S/12\). Use parallel spin frames at the endpoints. The contracted derivative of the spin transporter is \(-\gamma^i S_{ij}z^j/4\), obtained by contracting \(z^\nu R_{\alpha\beta\nu\mu}\gamma^\mu\gamma^\alpha\gamma^\beta/8\). The Van Vleck derivative adds \(+\gamma^i S_{ij}z^j/6\). Consequently the spinor propagator is
\begin{align*}
 {\cal S}(t,z)={}&2bc_b Z^{-b-1}\gamma^\mu z_\mu
    \left(1+\frac{S_{zz}}{12}-\frac{S Z}{48b}\right)\\
 &+\frac{c_b}{12} Z^{-b}\gamma^i S_{ij}z^j.
\end{align*}
The four-component trace of \(\gamma^0{\cal S}\gamma^0{\cal S}\) uses the flat numerator \(4(2bc_b)^2(t^2-\ell^2)\) when the reversal sign of the second propagator is factored out. The two spin-connection cross terms give \(-8(2bc_b)(c_b/12)S_{zz} Z^{-2b-1}\), and the scalar endomorphism multiplies the flat numerator by \(-SZ/(24b)\). The fermion determinant with \(D=\nabla+\ii gA\) gives a sign opposite to the flat gluon ratio. Using Eq.~\eqref{eq:moments} yields trace-free relative term \((1/6+d/(24b^2))S_{zz}\). Substituting \(S_{zz}=(d-1)K\ell^2\), \(S=d(d-1)K\) and including the endomorphism term gives
\begin{equation}\label{eq:rf}
 R_f=\frac{2b^2+1}{3(4b-1)}
\end{equation}
for the \(K\ell^2\) coefficient before attachment. This scalar term matters even in the final four-dimensional hyperbolic coefficient.

An independent trace check is available from conformal transport of the free conserved fermion current to $\mathbb R\times S^d_{1/q}$. The temporal component, up to the common normalization, is
\begin{equation*}
 \frac{2q^{2d}[1-\cos(q\ell)\cosh(qt)]}{[2(\cosh(qt)-\cos(q\ell))]^{d+1}} .
\end{equation*}
At $\ell=1$, the $q^2$ numerator after factoring $(1+t^2)^{-d-1}$ is $[-(t^4+1-6t^2)+(d+1)(1-t^2)^2]/12$. The ratios of the second and fourth moments to the zeroth moment are $1/(2d-1)$ and $3/[(2d-1)(2d-3)]$. Dividing by the flat moment of $1-t^2$ gives $(d^2-2d+3)/[6(2d-3)]$, precisely Eq.~\eqref{eq:rf} in continued dimension. Equation~\eqref{eq:attachTr} then gives $\rho_f^{\rm tr}=5\epsilon/3+O(\epsilon^2)$, so the flat fermion pole produces the finite trace coefficient $-20/9$. This checks the evanescent trace term independently of the spin-transporter calculation.

Attach the loop kernels and normalize by \(X_a g_B^2 k_1\ell^{2-d}\). With a further factor \(\as/(4\pi)\) taken out, the flat coefficients per \(\CA\) or \(\Tf n_f\) are denoted \(f_g,f_f\). For instance
\begin{align*}
 f_g&=16\pi^2(\mu^2 e^{\gamma_E}/4\pi)^\epsilon\,
      \frac{-c_b^2 A_b j_d\, k_{2-b}}{k_{-b} k_1}\,\ell^{2\epsilon},\\
 \frac{f_f}{f_g}&=-\frac{8b}{3d+1}.
\end{align*}
The trace-free \(\rho_g,\rho_f\) multiplying \(f_g,f_f\) are respectively
\((1/12-B_b/A_b){\cal A}(b)\) and \((1/12+d/(24b^2)){\cal A}(b)\). For the trace projection use Eq.~\eqref{eq:attachTr} with Eqs.~\eqref{eq:tracebubble}--\eqref{eq:rf}. Their first terms near \(b=1-\epsilon\) are
\[
 \begin{array}{c|cc}
 &P &K\\ \hline
 \rho_g &-3/10+(497/600)\epsilon & (7/15)\epsilon\\
 \rho_f &-(5/48)\epsilon & (5/3)\epsilon .
 \end{array}
\]
Each entry multiplies the indicated curvature and \(\ell^2\) in the normalized potential. The \(1/(b-1)\) in \(B_b\) explains why the STF gluon ratio has a nonzero leading term despite the zero of $\mathcal A(b)$.

\subsubsection{Connected two-source web and both color channels}
Use the two-line color operator \(X=T_1^a T_2^a\) before projecting any external state. The antiquark has the conjugate generator. At the order in metric jets used here the same-source temporal propagators agree, \(A(t)=H(t,0)\), and the opposite-source one is \(B(t)=H(t,\ell)\). Subtract tree iteration in the logarithm or in the inverse pair kernel. For two cross-source exchanges the connected crossed ordering has color \(C_A X\), and the connected nested ordering \(2 C_A X\). For example each nested color product differs from \(X^2\) by \(T_1^aT_1^b[T_2^b,T_2^a]=C_A X/2\). The four choices of endpoints produce \(2C_A X\). In the crossed ordering only two choices have the reversed generator ordering on one line. A mixed crossed diagram with the same-source kernel on either pair has \(-C_A X\). When the same-source pair is outer in the nested diagram its color is \(-2C_A X\); with the opposite-source pair outer its connected color is zero. These identities follow also from \(T_i^aT_i^bT_i^a=(C_F-C_A/2)T_i^b\). Purely isolated terms are removed. All connected color factors are therefore proportional to \(X\), an operator statement that covers the octet as well as the singlet.

The octet coefficient is defined from the normalized two-line amplitude with the endpoint color indices transported to the midpoint and projected onto the adjoint. Its low-energy adjoint time transporter belongs to $D_\tau$ in Eq.~\eqref{eq:action}, not to a third all-mode static source. In the matching, the corresponding point-adjoint propagation and endpoint overlaps are assigned to the low-energy theory: one extracts the separation-dependent extensive coefficient after dividing out isolated-line factors and subtracting retained low-energy iterations in the common regulator. This is a covariant projected coefficient; a standalone globally colored static energy is not assumed. In particular the color identity above pertains to two-line connected webs, not to a complete adjoint-source or detector matching.

Both orderings reduce to
\begin{equation}\label{eq:web}
\begin{aligned}
 V_{\rm web}=2g_B^4 C_A X_a\biggl[&\left(\int_0^\infty\dd t\,tA(t)\right)\left(\int_0^\infty\dd s\,B(s)\right)\\
 &-\int_{0<s<t}\dd s\,\dd t\,tB(t)B(s)\biggr].
\end{aligned}
\end{equation}
For example the two opposite-source orderings have weights \(s\) and \(t-s\) after arranging the time lengths with \(s<t\). Their sum gives \(t\). The same-source time moment in the first term is scaleless all-hard, but is retained for the regulated overlap. For the second term the needed integral at \(\ell=1\) is
\begin{align*}
 I(b,c)&=\int_0^\infty\dd v\,(1+v^2)^{-b}\int_v^\infty\dd t\,t(1+t^2)^{-c}\\
 &=\frac{\sqrt\pi\Gamma(b+c-3/2)}{4(c-1)\Gamma(b+c-1)}.
\end{align*}
The formula is continued from its convergent domain. In the trace-free part \(U^2/U=1+P\ell^2/12\). In the trace part the extra curvature after tree division is \(K\ell^2 \rho_W\), with
\begin{align*}
 \rho_W&=2A+B\,\frac{I(b-1,b)+I(b,b-1)}{I(b,b)}-q_1\\
 &=-\frac13+\frac83\epsilon+O(\epsilon^2).
\end{align*}
Here there are two shifted integrals because either temporal exchange carries the scalar endomorphism.

The web formula yields \(f_W=-32\pi^2(\mu^2e^{\gamma_E}/4\pi)^\epsilon c_b^2 I(b,b)\ell^{2\epsilon}/k_1\) per \(\CA\). For \(L_\ell=\log(\mu\ell)+\gamma_E\), expansion of the normalized flat factors gives
\begin{align}
 f_g&=\frac{5}{3\epsilon}+\frac{31}{9}+\frac{10}{3}L_\ell,\nonumber\\
 f_f&=-\frac{4}{3\epsilon}-\frac{20}{9}-\frac{8}{3}L_\ell,\qquad
 f_W=\frac{2}{\epsilon}+4L_\ell .\label{eq:flat}
\end{align}
These reproduce the flat coefficients \(a_1=(31C_A-20T_F n_f)/9\), \(\beta_0=(11C_A-4T_F n_f)/3\). There is no bulk vertex with three temporal legs in the ultrastatic optical calculation. Hard graphs tied to the spatial spokes contribute the overlaps, as discussed in Sec.~\ref{app:overlap}. The extensive connected terms in Eqs.~\eqref{eq:current} and \eqref{eq:web} have now accounted for a single polarization or two gauge exchanges at order \(g^4\).

\subsection{Geometric matching coefficients}
After the flat renormalization, the extra geometry before removing its infrared pole is
\begin{equation}\label{eq:rawextra}
\begin{aligned}
 &P\left[\CA\left(-\frac1{3\epsilon}+\frac{25}{72}-\frac23 L_\ell\right)+\Tf n_f\frac5{36}\right]\\
 &+K\left[\CA\left(-\frac2{3\epsilon}+\frac{55}{9}-\frac43 L_\ell\right)-\Tf n_f\frac{20}{9}\right].
\end{aligned}
\end{equation}
Its overall factor in four dimensions is \(X_a\as^2 \ell/(4\pi)\). The \(P\) finite gluon term is
\((5/3)(497/600)-(3/10)(31/9)=25/72\), plus zero from the finite part of \(f_W/12\) other than the logarithm. The \(K\) term comes from \(7/9\) in the polarization and \(16/3\) in the web. Its pole comes from the web in this trace continuation. Since \(P+2K=2\U_{nn}\), the bare soft infrared pole is the opposite of the geometric ultrasoft pole in the text. The residual scale after subtracting the pole at \(\nu\) replaces \(L_\ell\) by \(\log(\nu\ell)+\gamma_E\) in Eq.~\eqref{eq:rawextra}. In the normalized bracket this follows by multiplying the counterterm by \((\mu/\nu)^{2\epsilon}\) when writing it with the soft-loop measure at \(\mu\). The beta-function logarithms in Eq.~\eqref{eq:flat} retain \(\mu\).

Restoration of the physical metric includes the Weyl Ward insertion $-\beta_0\int\sqrt{\widetilde g}\ln N\,\mathcal F^2/(32\pi^2)$, with the coupling absorbed in $\mathcal F$ \cite{collins,wilson}. It follows from multiplying the inverse-coupling pole by $N^{-2\epsilon}$. Use $-\widetilde\nabla_i(f\widetilde\nabla^i)=\{f,L\}/2+\widetilde\nabla^2 f/2$, $f=\ln N$, and $G_2/G_1=-r^2/2$. The endpoint mean of $f$ is $r^2(\T_{nn}-a_n^2)/8$, and $\widetilde\nabla^2 f=R_{uu}-2a^2$. Combining with the distance correction to $L_\ell$ gives the extra term $\beta_0 D_N$, $D_N=\T_{nn}/6-R_{uu}/2+11a^2/12$. After replacing the overlap logarithm in Eq.~\eqref{eq:rawextra} by $L_\nu=\ln(\nu r)+\gamma_E$, the physical matching result is
\begin{equation}\label{eq:staticmatch}
 \begin{aligned}V_a^{\rm geom}={}&X_a\as(\mu)r\sum_k Z_k\biggl[T_k+\frac{\as(\mu)}{4\pi}\\
 &\times\{T_k(a_1+2\beta_0L_\mu)+\CA G_k+\Tf n_f F_k\}\biggr].\end{aligned}
\end{equation}
\begin{table*}[t]\centering
\caption{Static geometric coefficients in the optical subtraction. The last column is the shift in $G_k$ for the fixed-projection physical continuation; it is accompanied by the opposite ultrasoft shift.}\label{tab:static}
\begin{tabular}{lrrrr}\toprule
$Z_k$&$T_k$&$G_k$&$F_k$&$\delta G_k$\\\midrule
$R$&$-7/144$&$121/108-2L_\nu/9$&$-11/27$&$-1/6$\\
$R_{uu}$&$-5/16$&$55/12-4L_\nu/3$&$-5/3$&$-5/6$\\
$R_{nn}$&$-1/48$&$-11/36$&$1/9$&$1/6$\\
$C_{nunu}$&$5/24$&$47/36-4L_\nu/3$&$1/18$&$-1/3$\\
$a^2$&$3/8$&$-11/4+4L_\nu/3$&$1$&$1$\\
$a_n^2$&$-1/8$&$0$&$0$&$-1$\\\bottomrule
\end{tabular}
\end{table*}
In Ricci-flat freely falling geometry the result reduces to Ref.~\cite{wilson}. In a flat accelerated static completion the acceleration terms remain. These coefficients refer to the prescribed contours, not absolute energies of supported or freely falling sources. The independent canonical Coulomb-gauge calculation is given in Appendix~\ref{app:coulomb}.
\subsection{Common infrared overlap}\label{app:overlap}
Use a matching separator \(\kappa\) with \(r^{-1}\gg\kappa\gg \Delta,\sqrt{|{\rm Riem}|},|a|,\Lambda_{\rm QCD}\). A useful auxiliary implementation compares two static infrared completions whose low-frequency scalar and one-form propagators respect the same de Rham intertwining, and whose relevant connected hard amplitudes decay at time separations of order \(\kappa^{-1}\). Use low-frequency changes that agree with the local ultraviolet parametrix. The proof below uses their smooth differences. For homogeneous local kernels, the corresponding moments are taken first in their convergence domain and then continued. The cutoff on the linked spectral reference in Sec.~\ref{app:scheme} is an alternative explicit conversion at the same order.

Denote the propagator changes at the midpoint by
\begin{align*}
 {\cal C}_{ij}&=\delta D_{ij'}(0,0),\qquad {\cal C}_{00}=\delta H(0,0),\\
 {\cal M}_{ij}&=\int_0^\infty\dd t\, t\,\delta H_{;ij'}(t,0).
\end{align*}
Consider terms with one hard and one soft propagator, retaining the leading nonanalytic geometric term in \(r\) and suppressing extra powers of \(\kappa r\). The induced Coulomb Green function in the ordinary Feynman gauge has hard Fourier kernel
\begin{equation}\label{eq:overpol}
 \delta G_{\rm pol}(\boldsymbol q)=g_B^2\CA
       \left[\frac{4{\cal C}_{ij}q_iq_j}{(\boldsymbol q^2)^3}
       +\frac{{\cal C}_{00}-\tr{\cal C}}{(\boldsymbol q^2)^2}\right] .
\end{equation}
The first term is the soft constant in \(D_{ij'}\) multiplying the hard \(H_{;ij'}\) in Eq.~\eqref{eq:current}. The \(DD\) contact gives \(+{\cal C}_{00}\), since \(\widetilde\nabla_x^i\widetilde\nabla_y^{j'}D_{ij'}=\partial_t^2 H+\delta\). The quartic term proportional to \((f^{abc}B^b A_i^c)^2\) gives \(+\tr{\cal C}\) in the inverse propagator and hence \(-\tr{\cal C}\) in its inverse correction. These signs are needed in Eq.~\eqref{eq:overpol}. For the flat fractional kernels \(G_{j,d}=k_j r^{2j-d}\),
\begin{align*}
 -\partial_i\partial_jG_{3,d}&=-\frac{G_{1,d}}8\left[r_ir_j+\frac{r^2\delta_{ij}}{4-d}\right],\\
 G_{2,d}&=-\frac{r^2G_{1,d}}{2(4-d)}.
\end{align*}
Thus the \(\tr{\cal C}\) cancels at arbitrary \(d\), leaving
\begin{align}\label{eq:overpolpos}
 \delta V_{\rm pol}&=-g_B^2 X_a\Delta_d
          \left[r^ir^j{\cal C}_{ij}+\frac{r^2{\cal C}_{00}}{4-d}\right],\\
 \Delta_d&=\CA g_B^2 G_{1,d}/2.\nonumber
\end{align}
The regulated \({\cal C}_{00}\) term is not set to zero.

Vary Eq.~\eqref{eq:web}. Smoothing the outer exchange produces
\(\delta A(t)-\delta B(t)=r^ir^j\delta H_{;ij'}(t,0)/2+O(r^4)\) for the average of the two same-source propagators. This follows by Taylor expansion at the symmetric endpoints. Multiplying by the integrated hard exchange \(G_{1,d}/2\) gives the contribution proportional to \({\cal M}_{ij}\). Smoothing the inner exchange gives the time-ordering moment
\[
 \int_0^\infty\dd t\,[A_{\rm flat}(t)-B_{\rm flat}(t)]\,t^2
 =\frac{r^2 G_{1,d}}{2(4-d)} .
\]
For a constant inner propagator the ordered wedge \(0<s<t\) gives \(t^2/2\) after combining both variations. The \(A_{\rm flat}\) moment is scaleless in the homogeneous short-distance prescription. The formula also follows from the moment with $m=1$ in Eq.~\eqref{eq:moments}. Combining terms therefore gives
\begin{align}
 \delta V_{\rm web}&=+g_B^2 X_a\Delta_d
       \left[r^ir^j{\cal M}_{ij}+\frac{r^2{\cal C}_{00}}{4-d}\right],\nonumber\\
 \delta V_{\rm pol}+\delta V_{\rm web}
       &=-g_B^2 X_a\Delta_d r^ir^j({\cal C}_{ij}-{\cal M}_{ij}).\label{eq:overlap}
\end{align}
The temporal contribution cancels only in the sum. Terms with both propagators smooth are analytic in the short separation and do not contribute the local \(g^4 r Z_k\) coefficient.

Since \(\int_0^\infty\dd t\,t\,\delta G^E_{ij}={\cal C}_{ij}-{\cal M}_{ij}\), Eq.~\eqref{eq:overlap} equals the term linear in detuning in the variation of the ultrasoft self-energy, with \(Y_a=-X_a\Delta_d\). The equality is between the soft infrared dependence and its overlap in the ultrasoft calculation, so the variation cancels in matching. The boundary term giving \({\cal C}_{ij}\) arises from twice integrating \(\partial_t^2\delta D\) with time weight \(t\). Its sign is positive. A secondary check inserts a common proper-time interval in the local temporal and Hodge propagators. The terms logarithmic in heat time \(p\) give
\begin{align*}
 {\cal C}_{ij}&=\frac{S\delta_{ij}/6-S_{ij}}{16\pi^2}\int\frac{\dd p}{p},\\
 {\cal M}_{ij}&=\frac{S\delta_{ij}/6-S_{ij}/3}{16\pi^2}\int\frac{\dd p}{p}.
\end{align*}
The factor \(\int_0^\infty\dd t\,t e^{-t^2/(4p)}=2p\) makes the second coefficient explicit. The surviving difference is \(-S_{ij}/(24\pi^2)\) times the same integral. Continuing its large-\(p\) part gives \(-r^{2\epsilon}/\epsilon\) for \(\int_{r^2}^\infty\dd p\,p^{-1+\epsilon}\), with the same residue as Eq.~\eqref{eq:rawextra}. These tests cover both \(P\) and \(K\), both channels, and the cutoff-dependent finite part of the retained overlap. Dressing or completing the geometry at large distances still changes the infrared correlator and hence can change physical energies.

\section{Ultrasoft contributions}\label{sec:us}
\subsection{Local electric singularity}
\label{app:param}
For the local Riemannian calculation use \(D=d+1=4-2\epsilon\), \(b=(D-2)/2=1-\epsilon\), \(c_b=\Gamma(b)/(4\pi^{b+1})\), and \(Z=2\sigma\), with \(\sigma\) Synge's world function. The vector wave operator in Feynman gauge has Euclidean representative \(-\nabla^2\delta_\alpha{}^\beta+R_\alpha{}^\beta\). Integration of its first two heat coefficients gives
\begin{equation}\label{eq:paramapp}
\begin{aligned}
 D_{\beta\delta'}={}&c_b\left[Z^{-b}\Delta_{\rm VV}^{1/2} g_{\beta\delta'}
        +\frac{Z^{1-b}}{4(b-1)}(a_1)_{\beta\delta'}\right]+\cdots,\\
 [a_1]_{\beta\delta}&=R\,\delta_{\beta\delta}/6-R_{\beta\delta}.
\end{aligned}
\end{equation}
Here \(\Delta_{\rm VV}\) is the Van Vleck determinant and \(g_{\beta\delta'}\) the parallel propagator. At \(b=1\) the constant infrared piece in the second term is smooth locally and the singular part is \(-c_1 a_1\log Z/4\). Differentiation to field strengths must precede setting \(b=1\). This construction uses the usual covariant vector Hadamard coefficients and Ward identity \cite{frob,vass}.

Transport both frames to the geodesic and let \(v\) be its Riemannian unit tangent, with separation \(t\). The first variation and Jacobi equation give
\begin{align*}
 Z&=t^2,\quad Z_{;A}=2t v_A,\quad Z_{;C'}=-2t v_C,\\
 Z_{;AC'}&=-2\delta_{AC}-\frac{t^2}{3}R^E_{AvCv},\\
 \Delta_{\rm VV}^{1/2}&=1+t^2 R^E_{vv}/12,\\
 (\Delta_{\rm VV}^{1/2})_{;A}&=tR^E_{Av}/6,\\
 (\Delta_{\rm VV}^{1/2})_{;C'}&=-t R^E_{Cv}/6,\\
 (\Delta_{\rm VV}^{1/2})_{;AC'}&=-R^E_{AC}/6.
\end{align*}
Only terms contributing through curvature divided by \(t^{2b}\) are written. The parallel-propagator derivatives can be found in radial gauge for the frame connection, \(A_A(y)=y^B R^E_{BA}/2+O(y^2)\) as a matrix on frame indices. The transporter for the chord is \(\mathbf 1-R^E_{BA}y'^B y^A/2+O(y^3)\). Covariant endpoint differentiation then gives
\begin{align*}
 g_{BD';A}&=-tR^E_{BDAv}/2,\quad g_{BD';C'}=-tR^E_{BDCv}/2,\\
 g_{BD';AC'}&=R^E_{BDAC}/2.
\end{align*}
These last derivatives contribute even though the undifferentiated parallel propagator on the central segment is the identity in aligned frames.

Antisymmetrize the vector derivative in each electric index pair. The following entries are the contributions to the \emph{ordinary} Euclidean \(F_{iv}F_{jv'}\), with \(c_b t^{-2b}\) factored out.
\begin{center}\footnotesize
\setlength{\tabcolsep}{1.5pt}
\begin{tabular}{@{}lrrrr@{}}
\toprule
Term & \(R^E_{ivjv}\) & \(R^E_{ij}\) & \(R^E_{vv}\delta_{ij}\) & \(R^E\delta_{ij}\)\\
\midrule
Scalar & \(b/3\) & \(-1/6\) & \(b/6-(2b-1)(b-1)/6\) & \(0\)\\
Transport & \(-(2b-1)\) & \(0\) & \(0\) & \(0\)\\
\(a_1\) & \(0\) & \((2b-1)/2\) & \(-1/2\) & \((1-b)/6\)\\
\bottomrule
\end{tabular}
\end{center}
For example the first line comes from the mixed derivative of \(Z^{-b}\) and the cross terms with the first derivatives of \(\Delta_{\rm VV}^{1/2}\). In the last line the factor \(b-1\) from differentiating \(Z^{1-b}\) cancels its denominator. The \(\delta_{ij}\) and \(vv\) contractions of \((\log Z)_{;AC'}\) would give the same four-dimensional limit. The flat contribution to the physical \(G^E\) is \(4b^2c_b\,\delta_{ij}t^{-2b-2}\).

Rotate \(R^E_{ivjv}\to-\T_{ij}\) and \(R^E_{vv}\to-R_{uu}\) and apply the electric sign. The rotation is a calculation on local jets, so it gives the Hadamard coefficients for a smooth Lorentzian metric without requiring a global analytic continuation. Denote the dimensionally continued physical curvature coefficient divided by \(c_b\) by \(f^{\rm phys}_{ij}\). The geodesic part is
\begin{equation}\label{eq:physicalDcurv}
\begin{aligned}
 f_{ij}^{\rm phys,curv}={}&(1-5b/3)\T_{ij}-(b-2/3)R_{ij}\\
 &-\frac{b^2-2b+2}{3}R_{uu}\delta_{ij}+\frac{b-1}{6}R\delta_{ij}.
\end{aligned}
\end{equation}
In particular the dimension-four expression is
\(-[2\T_{ij}+R_{ij}+R_{uu}\delta_{ij}]/(12\pi^2)\) after restoring \(c_1=1/(4\pi^2)\).

\subsubsection{Acceleration and the general result}
At weight two the pure curvature result can be added to the flat accelerated calculation. There is no term of mixed weight \(aR\) at this order. In the flat physical continuation a uniformly accelerated line becomes a Euclidean circle. Put \(t_a=2\sin(at/2)/a\), \(\chi=at/2\), and use the Fermi--Walker frames. The electric component in the acceleration plane is \(4b^2c_b t_a^{-2b-2}\). A perpendicular component is \(4bc_b(b\cos^2\chi+\sin^2\chi)t_a^{-2b-2}\). To see the second formula, the electric bivector is a sum of a time--space and a space--space bivector. At symmetric endpoints their angles are opposite. The two elementary flat contractions are \(-4b^2 c_b t_a^{-2b-2}\) and \(4b c_b t_a^{-2b-2}\) before the electric sign, and the mixed contraction vanishes. Expanding the two results supplies the remainder of Eq.~\eqref{eq:physicalDcurv},
\begin{equation}\label{eq:physicalDacc}
 f_{ij}^{\rm phys,acc}=\frac{b(b^2-2b+3)}{3}a^2\delta_{ij}
                           +b(b-1)a_i a_j .
\end{equation}
The physical calculation keeps the acceleration anisotropy in \(D\) dimensions even though it vanishes in four. Independent weight-two terms involving the jerk have already been excluded by the tensor symmetries.

Substituting the four-dimensional Ricci decomposition of $\T_{ij}$ gives Eq.~\eqref{eq:U}. The resulting short-distance electric tensor and Euclidean correlator are
\begin{align}
 \U_{ij}&=C_{i0j0}+(R_{uu}+R/6-a^2)\delta_{ij},\label{eq:U}\\
 G^E_{ij}(t)&=\frac{\delta_{ij}}{\pi^2t^4}-\frac{\U_{ij}}{6\pi^2t^2}+\cdots .\label{eq:GE}
\end{align}
The local Wightman boundary value has $(s-\ii0)^{-4}$ and $+\U_{ij}(s-\ii0)^{-2}/(6\pi^2)$. Equation~\eqref{eq:GE} specifies a singularity, not a globally positive bath spectrum. In a static optical frame $N^{-2}\widetilde R_{ij}=2\U_{ij}$ at this order. For an FLRW conformal vacuum $\U_{ij}=(-H^2-2\dot H+k/A^2)\delta_{ij}$, with scale factor $A(t)$ and spatial-curvature label $k$. In flat uniformly accelerated motion $\U=-a^2\one$. These limits check the short-time free correlator. They do not imply thermal equivalence of the physical states.
For an independent spatially flat FLRW check, let $s$ be the proper-time lag and use a conformal vacuum. The electric Wightman function is proportional to $[A(t+s/2)A(t-s/2)]^{-2}[\eta(t+s/2)-\eta(t-s/2)-\ii0]^{-4}$, with conformal time $\eta$. The two midpoint expansions give $A_+A_-=A^2(1+\dot Hs^2/4+\cdots)$ and $\Delta\eta=(s/A)[1+(H^2-\dot H)s^2/24+\cdots]$. Thus the normalized relative correction is $-(H^2+2\dot H)s^2/6$, recovering the stated $\U$ and its Lorentzian sign.
\subsection{Static and mobile renormalization}
The color weights of the dipole loop on channel $a$ with intermediate $c$ are
\begin{equation}
 \begin{gathered}w_{so}=\CF,\quad w_{os}=\Tf/N_c,\\w_{oo}=(N_c^2-4)/(4N_c),\quad w_{ss}=0.\end{gathered}
\end{equation}
The gap $\Delta=\CA\as/(2r)$ obeys $\sum_c w_{ac}(V_c-V_a)=-X_a\Delta$. The geometric pole is $X_a\CA\as^2 r\U_{nn}/(6\pi\epsilon)$ on shell. Indeed the detuning integral is $\Gamma(-1+2\epsilon)(V_c-E-\ii0)^{1-2\epsilon}$. The corresponding running is
\begin{equation}\label{eq:pinnedRG}
 \nu\partial_\nu V_a^{\rm geom}=-\frac{X_a\CA\as^2 r}{3\pi}\U_{nn}.
\end{equation}
The $\nu$ derivative of Table~\ref{tab:static} gives precisely Eq.~\eqref{eq:pinnedRG}. Ordinary running cancels the $L_\mu$ terms at this order. The octet-to-octet kinetic pole at zero static gap is an EOM operator and changes the source normalization, not the on-shell static potential.

For mobile relative dipoles write $A_a=h_a-\ii D_\tau$ and $\mathcal L_{ca}B=A_cB-BA_a$. The local pole in the convention $\ii D_\tau-h_a-\Sigma_a$ is
\begin{equation}\label{eq:pole}
 \Sigma_a^{\rm UV}=\frac{g^2}{12\pi^2\epsilon}\sum_c w_{ac}\Herm[r_i A_c^3 r_i-\U_{ij}r_i A_c r_j].
\end{equation}
This follows by covariantly Taylor expanding $U_c(t,t-s)(r_i\psi_a)(t-s)$ with the time-ordered $U_c$, not by commuting the Hamiltonians at different times. Expand $A_c^m r_i=\sum_{j=0}^m\binom mj(\mathcal L_{ca}^{m-j}r_i)A_a^j$. Terms with $j>0$ are local inverse-propagator operators; their removal changes endpoint sources. The linear tensor moment is
\begin{align}
 \Herm(r_i(h_c-E)r_j)_{(ij)}\nonumber\\={}&\tfrac12\{r_ir_j,h_a-E\}\nonumber\\
 &+\tfrac12[r_i,[h_a,r_j]]\nonumber\\
 &+\tfrac12[r_i(h_c-h_a)r_j\nonumber\\
 &\qquad+r_j(h_c-h_a)r_i].
\end{align}
At central rest with leading $p^2/M+V_a$, its contraction with $\U$ gives $\tr\U/M+(V_c-V_a)\U_{rr}$ after EOM removal. In particular the same-channel mobile contribution need not vanish. With $z_c=\Delta V_{ca}$, $K_a=\ii D_\tau-h_a$ and $D_{ij}^{ac}=\delta_{ij}/M+z_c r_i r_j$, the linear-kernel pole is
\begin{equation}
\Sigma_{a,\U}^{\rm UV}=-\frac{g^2}{12\pi^2\epsilon}\sum_c w_{ac}\left[\U_{ij}D_{ij}^{ac}-\frac12\{\U_{ij}r_i r_j,K_a\}\right].
\end{equation}
Here the geometric tensor is evaluated on the central line, with midpoint Hermitian ordering. If it varies over the center support, the additional commutator $[\U_{ij},[h_a,r_ir_j]]/4$ in the contracted moment is derivative and mixed-recoil suppressed in this projection. The pole is canceled by the inverse-kernel counterterm $\{\delta Z_a,K_a\}/2-\delta h_a$ with $\delta Z_a=g^2\sum_c w_{ac}\U_{ij}r_i r_j/(12\pi^2\epsilon)$ and $\delta h_a=g^2\sum_c w_{ac}\U_{ij}D_{ij}^{ac}/(12\pi^2\epsilon)$. Both source legs are transformed with the inverse-field normalization. In the cubic term the binomial identity gives the remaining EOM polynomials explicitly: $\Herm\sum_{j=1}^3\binom{3}{j}r_i(\mathcal L_{ca}^{3-j}r_i)A_a^j$. They are local and removed order by order with the same endpoint transformation; this does not remove the Hermitian $j=0$ term.

The curvature dependence of Eq.~\eqref{eq:pole} is not exhausted by the electric singularity. Although the common tide $F$ cancels in the potential difference, it does not commute with the kinetic descendants. Put $z_c=\Delta V_{ca}=V_c-V_a$ and $f_i=2\partial_i F/M$. At linear $F$,
\begin{equation}\label{eq:cubic}
 \delta_F\Herm(r_i\mathcal L_{ca}^3r_i)=\frac{\partial_i f_i}{M}+3z_c r_i f_i+r^2 f_i\partial_i z_c .
\end{equation}
With $f_i=\T_{ij}r_j$ and Coulomb $r\partial_r z_c=-z_c$ this is $\tr\T/M+2z_c\T_{rr}$. To verify this identity, write $\mathcal L_0 r_i=z_c r_i-2\partial_i/M$ in a local Cartesian frame. Then $[F,\mathcal L_0r_i]=f_i$ and $[F,\mathcal L_0^2r_i]=r_i f_j\partial_jz_c+2z_cf_i$. The other term is $\mathcal L_0f_i=[p^2/M,f_i]+z_cf_i-\ii\dot f_i$. The Hermitian outer product removes the last term, and $\Herm(r_i[p^2/M,f_i])=\partial_i f_i/M$. This yields Eq.~\eqref{eq:cubic} without using a static gap in place of the propagator. In the canonical relative-dipole, central-rest Hamiltonian representation the leading geometric running is therefore
\begin{equation}\label{eq:mobileRG}
 \begin{aligned}\nu\partial_\nu h_a^{\rm geom}={}&\frac{2\as}{3\pi}\sum_c w_{ac}\biggl[\frac{\tr(\U-\T)}{M}\\
 &+z_c(\U-2\T)_{ij}r^ir^j\biggr].\end{aligned}
\end{equation}
As a check, the known flat-space on-shell cubic moment in Eqs.~(32)--(33) of Ref.~\cite{pineda2011} can be varied by an isotropic common potential $F=Mt r^2/4$. With reduced mass $m_r=M/2$ its variation is $\nabla^2F/(2m_r^2)+2z_c rF'/m_r=3t/M+2z_c tr^2$, in agreement with Eq.~\eqref{eq:cubic}. The flat cubic moment is not new; the tensor variation and its combination with the geometric electric correlator give Eq.~\eqref{eq:mobileRG}.
\subsubsection{Closure of the retained pole and endpoint terms}
At this order the singularity of the electric two-point function can be expanded in transported frames at the midpoint of the two proper times. The symmetric Hadamard singular part is invariant under joint interchange of the two indices and the two endpoints. A term of one higher geometric weight, with a single inverse power of the lag, is therefore antisymmetric in the indices in the midpoint expansion. Its possible logarithmic residue multiplies $r_i r_j$ at zero lag and vanishes. This statement uses only the leading commuting electric dipoles; an independent spin or higher-multipole vertex need not obey it. On expanding instead at the later endpoint, the symmetric coefficient has the Taylor term $-s\dot\U_{ij}/2$. With $B=\U_{ij}r_i r_j$, the resulting derivative term is precisely the symmetric ordering $\{B,A_a\}/2=BA_a-\ii\dot B/2+[h_a,B]/2$. Thus it belongs to the inverse-propagator term, rather than a new Hermitian multiplication potential. The derivative of the common tide in Eq.~\eqref{eq:cubic} is likewise proportional to $-\ii r_i\dot f_i$, whose Hermitian part vanishes.

For the cubic pole put $Q_j=\binom{3}{j}r_i\mathcal L_{ca}^{3-j}r_i$. Each term with $j>0$ has the exact form
\begin{equation}
 \Herm(Q_j A_a^j)=B_jA_a+A_a B_j^\dagger,\quad B_j=Q_j A_a^{j-1}/2.
\end{equation}
At first loop order it is a local congruence of the inverse propagator. The corresponding source transformation is required on both ends; temporal integrations by parts at finite endpoints are not discarded as physical terms. This removes the redundant derivative polynomial, without invoking any inverse singlet--octet gap. A normalized observable must be transformed by the same congruence. On the central line the $j=0$ Hermitian multiplication terms at the retained order are the linear moment and the common-force variation already evaluated. For example, at fixed Coulombic scale the ordinary cubic self-energy counts as $\as r^2E_B^3$. Changing the geometric kinetic term, static potential or dipole by $O(Rr^2)$ gives $O(\as v^2 R/M)$, while $F/E_B=O(\|\T\|/E_B^2)$ gives $O(\as\|\T\|/M)$. The former effects are beyond the latter accuracy. Magnetic/Pauli vertices, higher electric multipoles and the shift terms in the Fermi Hamiltonian are velocity or gradient suppressed in this counting; an electric dipole may be replaced by a Pauli dipole only at cost $1/(Mr)=v$. A radiative correction to a leading vertex costs another power of the coupling. Center derivatives and center velocities are not included in this rest projection. This establishes the local two-dipole closure at the retained geometric and velocity orders; it does not classify the entire time-dependent recoil basis or finite endpoint operators at higher orders.

Equation~\eqref{eq:mobileRG} is the contribution of two relative electric dipoles in the central-rest representation, including the common enhanced mass tide. For $a=o$ it is the corresponding projected relative-dipole contribution, not the full renormalization of a mobile adjoint source: center-adjoint monopole propagation, its recoil, and independent source insertions have not been matched here. Such additional octet matching is not required for the leading singlet two-E1 loop, which uses the tree octet Hamiltonian. The local running does not fix the general finite-mass matching. In vacuum on a geodesic $\U=\T$ and the gap contribution has the opposite sign to Eq.~\eqref{eq:pinnedRG}. The opposite signs refer to Wilson coefficients in distinct matching problems, not to a reversal of a physical energy shift. In flat spatial FLRW, $\T=-(\dot H+H^2)\one$, so $\U-2\T=H^2\one$ and $\tr(\U-\T)=-3\dot H$. A vanishing geometric bath singularity does not in general eliminate the common-tide contribution. The finite mobile coefficients require separate matching. Indeed the difference between Eqs.~\eqref{eq:mobileRG} and \eqref{eq:pinnedRG} has the local derivative
\begin{equation}
 \begin{aligned}
 \nu\partial_\nu(h_a^{\rm geom}-V_a^{\rm geom})
 &=\frac{2\as}{3\pi}\sum_c w_{ac}\\[-1ex]
 &\quad\times\left[\frac{\tr(\U-\T)}{M}-2z_c\T_{rr}\right].
 \end{aligned}
 \label{eq:mobileboundary}
\end{equation}
Integration determines a logarithm but not its finite boundary value. In the homogeneous ultrastatic applications the trace term is a common scalar and the second term vanishes; in a general tide neither argument fixes the finite matching. In particular $[p^2/M,F]=-\ii\T_{ij}\{p_i,r_j\}/2$ stays finite at fixed curvature and separation as $M$ increases. Removing the relative kinetic operator before forming the cubic moment is not the same operation as taking this formal large-mass limit after the moment. Moreover, at fixed curvature and separation the mass tide grows with $M$; a bound-state limit also requires specifying how $a_0,E_B$, curvature and support vary.

\subsubsection{Finite changes of representation}
The finite multiplication coefficient in Eq.~\eqref{eq:mobileboundary} is specific to the Hamiltonian representation. For example, in the singlet sector at central rest in a stationary Ricci-flat tide, take the Hermitian generator
\begin{equation}
 G_\lambda=\tfrac12\{p_i,w_i\},\qquad
 w_i=\lambda\frac{\CA\as}{\pi}\T_{rr}r_i,
\end{equation}
where $\lambda$ is real and $\T_{rr}=\T_{jk}r^jr^k$. To first order in curvature the transformation $h'_s=e^{\ii G_\lambda}h_se^{-\ii G_\lambda}$ changes the leading Coulomb potential by
\begin{equation}\label{eq:unitarypotential}
 \ii[G_\lambda,V_s^{(0)}]
 =w_i\partial_i V_s^{(0)}
 =\lambda\frac{\CF\CA\as^2}{\pi}r\T_{nn}.
\end{equation}
At the same order the relative kinetic term transforms as
\begin{equation}\label{eq:unitarykinetic}
 \ii[G_\lambda,p^2/M]
 =-\frac{1}{2M}\{p_i,\{p_j,\partial_i w_j\}\}.
\end{equation}
These identities follow from $[p_i,f]=-\ii\partial_i f$ for a multiplication operator $f$. The commutator with the mass tide is quadratic in curvature. The two terms in Eqs.~\eqref{eq:unitarypotential} and \eqref{eq:unitarykinetic} have the same Coulombic order, $O(\as R/M)$ for $r\sim (M\as)^{-1}$. Their sum is a commutator with the leading flat Hamiltonian and has vanishing matrix elements within an unperturbed energy eigenspace. More generally, states, dipoles and probe operators must be transformed along with the Hamiltonian (with an additional time-derivative term in the Hamiltonian for a time-dependent transformation). The finite shift in Eq.~\eqref{eq:unitarypotential} is therefore not a separately measurable energy shift. This freedom does not supply the boundary value in the fixed canonical matching representation. Independent adjoint-source, recoil and gravitational probe coefficients remain necessary for more general dynamical applications.

\subsection{Ultrasoft reference, cutoff, and finite scheme conversion}\label{app:scheme}
In the optical continuation the coefficient of \(c_b t^{-2b}\) contracted with \(nn\) is
\begin{equation}\label{eq:optf}
 f_{\mathrm{opt}}=-(b-2/3)P-\frac{b(3-b)(2b-1)}{3}K .
\end{equation}
Relative to their four-dimensional values its \(P,K\) parts are \(1-3\epsilon\) and \(1-5\epsilon/2\). Define $b^E_P=-P/(12\pi^2)$ and $b^E_K=-K/(6\pi^2)$, the two terms in the four-dimensional $t^{-2}$ electric coefficient, and $\mathcal B_{ac,T}=g^2r^2w_{ac}b^E_T$ for $T=P,K$. At fixed detuning in the tree-normalized bracket the reference has the Laurent form
\begin{align}\label{eq:reference}
 \Sigma^{\rm ref}_{ac,T}&=-{\cal B}_{ac,T} z_c
 \left[-\frac1{2\epsilon}+\log\frac{2(z_c-\ii0)}{\nu}+c_T\right],\\
 c_P&=\frac12,\qquad c_K=\frac14.\nonumber
\end{align}
Here \(z_c=V_c-E\) is real before the indicated prescription. To derive the constants, multiply \(\Gamma(-1+2\epsilon)\) by
\((\nu/2)^{2\epsilon}e^{\gamma_E\epsilon}\Gamma(1-\epsilon)(1-p_T\epsilon)z_c^{-2\epsilon}\),
with \(p_P=3,\ p_K=5/2\). The finite constant is \(p_T/2-1\). When using the continued tree gap as in Eq.~\eqref{eq:overlap}, its overall factor is divided out with the same tree factor as the soft pole. Terms of order \(\epsilon\) from a change to a dimensionful subtraction must be retained. The negative octet detuning is analytic continuation with the lower-half-plane logarithm. The resulting local reference is a subtraction kernel. A physical cut or level shift also depends on the full state and the chosen gauge-invariant probe for the octet.

With \(\Sigma_{\rm ref}^{\rm ren}\) defined by removing the pole in Eq.~\eqref{eq:reference}, the gauge-linked leading spectral weight at four dimensions gives an explicit cutoff reference
\begin{equation}\label{eq:cut}
 \Sigma^<_{ac,T}(\kappa)=-{\cal B}_{ac,T}
 \left[\kappa-z_c\log\frac{\kappa+z_c-\ii0}{z_c-\ii0}\right] .
\end{equation}
Both the power overlap and the logarithmic overlap must be subtracted. In the corresponding dimensionally continued low-frequency expansion use \(\omega/(\omega+z_c)=1-z_c/\omega+\cdots\) with weight
\(\omega^{-2\epsilon}(\nu/2)^{2\epsilon}[1+(2-p_T)\epsilon]\). Integrating the homogeneous terms up to \(\kappa\) gives
\[
 \Sigma^{\rm ov}_{ac,T}
 =-{\cal B}_{ac,T}\kappa-{\cal B}_{ac,T}z_c
    \left[\frac1{2\epsilon}+\log\frac{\nu}{2\kappa}-c_T\right].
\]
Its pole is opposite the ultrasoft ultraviolet pole. The expansion is only a continuation of overlap terms. At \(\omega=0\) one uses the unexpanded integral for the reference. For each channel,
\[
 V_a^\kappa=V_a^{\rm opt}
      +\sum_{c,T}(\Sigma^{\rm ref,ren}_{ac,T}-\Sigma^<_{ac,T}(\kappa)).
\]
Then \(V_a^\kappa+\sum\Sigma^<\) is independent of the separator. In a cutoff description the power shift is analytic in \(r\) and is carried with the same multipole subtraction. The \(oo\) contribution at \(z_o=0\) is of precisely this type. The spectral separator tests factorization of the linked correlator. It need not act on separate gauge-potential components of the full theory.

\subsubsection{Combined mobile overlap in vacuum}
For the central-rest singlet on a geodesic in vacuum, $\U=\T$ and $\tr\T=0$. The cancellation of power overlaps can be examined directly in four dimensions. Take a stationary local reference, write $z=\CA\as/(2r)>0$, $d_\omega=\omega+z$, and $K_s=2\CF\as/(3\pi)$. After the inverse-propagator reduction the resolvent acts on the dipole as $(\omega+\mathcal L_{os})^{-1}r_i$. In a separated-frequency expansion, retain the leading geometric electric term and the first kinetic descendant enhanced by the common mass tide. With $\mathcal L_0 B=[p^2/M,B]+zB$ for multiplication operators, the first variation of the resolvent is $- (\omega+\mathcal L_0)^{-1}[F,\,\cdot\,](\omega+\mathcal L_0)^{-1}$, where on differential operators $\mathcal L_0$ is the full flat singlet--octet Liouvillian. Since $[F,r_i/d_\omega]=0$, the first nonzero term follows from the kinetic insertion and $[F,[p^2/M,b]]=f_j\partial_j b$ for multiplication operators $b$, with $f_j=\T_{jk}r_k$. At the retained recoil order this gives
\begin{align}
 \delta_F(\omega+\mathcal L_{os})^{-1}r_i
 &=\frac{1}{d_\omega^2}f_j\partial_j\frac{r_i}{d_\omega}\nonumber\\
 &=\frac{f_i}{d_\omega^3}
 -\frac{r_i f_j\partial_j z}{d_\omega^4}.
 \label{eq:mobresolvent}
\end{align}
Here the multiplication terms of the expansion are used in the outer resolvent; additional kinetic insertions are beyond this separated-frequency projection. Contracting with $r_i$ and using $f_j\partial_j z=-z\T_{nn}$ yields $\T_{rr}(d_\omega^{-3}+zd_\omega^{-4})$. The spectral weights of the $t^{-4}$ and $-\U_{ij}t^{-2}/6$ terms in Eq.~\eqref{eq:GE}, including the common self-energy sign, then give the combined reference over a frequency interval $[\kappa,\Lambda]$:
\begin{align}
 \Sigma_{\rm mob}^{[\kappa,\Lambda]}
 &=K_s\T_{rr}\int_\kappa^\Lambda\dd\omega\,\mathcal I(\omega,z),\\
 \mathcal I(\omega,z)&=\frac{\omega}{d_\omega}
 -\frac{\omega^3}{d_\omega^3}
 -\frac{z\omega^3}{d_\omega^4}\nonumber\\
 &=\frac{z}{d_\omega}-\frac{2z^3}{d_\omega^3}
 +\frac{z^4}{d_\omega^4}.
 \label{eq:moboverlap}
\end{align}
The constant terms cancel between the electric singularity and the mass-tide variation. The large-frequency overlap is $\mathcal I=z/\omega+O(z^2/\omega^2)$, with no linear power overlap at this order. Its integral is
\begin{equation}\label{eq:mobcut}
 \Sigma_{\rm mob}^{[\kappa,\Lambda]}=K_s\T_{rr}
 \left[z\ln\frac{d_\omega}{z}+\frac{z^3}{d_\omega^2}
 -\frac{z^4}{3d_\omega^3}\right]_\kappa^\Lambda.
\end{equation}
The logarithmic cutoff coefficient is $+K_s z\T_{rr}$, opposite to the pinned electric reference and canceled by the running $-K_s z\T_{rr}$ of Eq.~\eqref{eq:mobileRG} in a cutoff factorization. This comparison uses a separator above the resolved energies and below the eliminated soft scale; the large-frequency expansion tests the homogeneous overlap rather than extending the EFT to arbitrarily large physical frequencies. The finite endpoint terms in Eq.~\eqref{eq:mobcut} are those of the four-dimensional reference alone. They do not determine the continued-operator finite subtraction, the state-dependent response or the microscopic mobile matching coefficient.

\subsubsection{Dimensional scheme conversion}
There is also a dimensional-scheme test with no numerical regulator. Define a physical continuation by holding \(\T_{nn},R_{nn},R_{uu},R,a^2,a_n^2\) fixed through the \(\epsilon\) expansion. The fixed Weyl slot is the four-dimensional algebraic combination \(\T_{nn}-(R_{uu}-R_{nn})/2-R/6\). Use the same normalized gap subtraction as in Eq.~\eqref{eq:reference}. In the tensor calculation of Sec.~\ref{app:param}, Eqs.~\eqref{eq:physicalDcurv} and \eqref{eq:physicalDacc} give
\begin{align}
 f_{\rm phys}-f_{\rm opt}&=\epsilon\,e(n)+O(\epsilon^2),\nonumber\\
 e(n)&=-\frac{C_{nunu}}3-\frac{5R_{uu}}6
              +\frac{R_{nn}}6-\frac R6+a^2-a_n^2 .\label{eq:e}
\end{align}
For example \(f_{\rm phys}\) has \(\epsilon\) coefficient
\(5\T_{nn}/3+R_{nn}-R/6-2a^2/3-a_n^2\), while \(f_{\rm opt}\) has \(P+5K/3\). Multiplying by the pole of the time integral changes the renormalized ultrasoft reference by
\[
 \Sigma^{\rm ref,ren,phys}_a-\Sigma^{\rm ref,ren,opt}_a
    =-X_a\frac{\CA\as^2r}{4\pi}\,e(n).
\]
The soft coefficient changes by the opposite amount, which is the last column of Table~\ref{tab:static}. In Rindler this gives the prior finite \(\CA(a^2-a_n^2)\) in the physical soft loop bracket. The change is generated by the difference of the evanescent free electric kernels. A pure rescaling of \(P\) by \(1+\eta\epsilon\) in the \emph{optical} scheme, with the tree convention held, instead shifts the soft bracket by \(-\CA\eta P/3\), and the ultrasoft part by the opposite. With \(P=2C_{nunu}\) this reproduces the optical deformation of Ref.~\cite{wilson}. A dimensionally continued physical Ricci-flat metric and a rescaled optical Ricci tensor describe distinct prescriptions. Finite shifts of the tree normalization require a corresponding adjustment to these formulas.

For a mobile state the same evanescent term gives
\begin{align*}
 \delta_{\rm sch}\Sigma^{\rm ren}_a&=\frac{g^2}{8\pi^2}\sum_c w_{ac}\Herm[e_{ij}r_i(h_c-E)r_j],\\
 e(n)&=e_{ij}n_i n_j.
\end{align*}
The signed dipole identity separates the gap part (opposite the static shift), a common trace scalar, and an inverse-propagator overlap. On $H^3$ one has $e_{ij}=(2q^2/3)\delta_{ij}$, so the scalar is $+\as\CF q^2/(\pi M)$ in the singlet and cancels from the level difference. The finite gap contribution to the $2S-2P$ difference is $+\CA\as q^2/(3\pi M)$, canceled by the soft shift. On an ultrastatic slice the evanescent tensor is isotropic and does not affect the magnetic-sublevel difference. This argument concerns the stated stationary projections, not absolute finite mobile matching.

The log derivatives in Eqs.~\eqref{eq:reference} and \eqref{eq:staticmatch} cancel for both channels by Eq.~\eqref{eq:pinnedRG}. Independently the explicit \(2\beta_0 L_\mu\) cancels the leading \(\mu\)-running of \(\alpha_s(\mu)\) multiplying the full tree tensor. State-dependent remainders added to the ultrasoft kernel are finite at this order. In a time-dependent calculation the corresponding local counterterms cancel the scale derivative of the Dyson kernel, with the same $K_a$-dependent renormalization and endpoint overlaps as in Eq.~\eqref{eq:pole}. This cancellation leaves the finite boundary value in Eq.~\eqref{eq:mobileboundary} to be matched for a general physical amplitude.
\subsection{Conformal-cylinder benchmark}\label{sec:cuspcheck}
An independent static check is available on $\mathbb R_t\times S^3_{1/q}$. Set $\beta_0=0$ formally at the order $\as^2$ (no all-orders conformal theory is assumed). Radial quantization maps a Euclidean cusp of angle $\phi=\pi-\delta$, $\delta=qr$, to a singlet static energy on the cylinder, $q\Gamma_{\rm cusp}$ up to a separation-independent term \cite{cusp,cuspreview}. This is a fixed-angle perturbative expansion with $\as\ll\delta\ll1$, hence $\Delta=\CA\as/(2r)\ll q$, different from the bound-state hierarchy in Sec.~\ref{sec:spectra}.

For completeness, write the order-$\as^2$ cusp coefficient in units $\CF\CA(\as/\pi)^2$ as $G=(\widetilde A_2+\widetilde A_3)/2+\widetilde A_1/3$, using Eqs.~(5.2)--(5.5) of Ref.~\cite{cusp}. Here tildes subtract the zero-angle limit. Analytic continuation of the two-loop functions gives, with $c=\cot\phi$, $I=\int_0^\phi u\cot u\,\dd u$ and $J=\int_0^\phi(\phi-u)u\cot u\,\dd u$,
\begin{align*}
 \widetilde A_1&=\phi c-1,\quad A_2=\pi^2/3-\phi^2-2c I,\\
 A_3&=-c(\pi^2\phi-\phi^3)/3+2c^2J .
\end{align*}
Putting $l=\ln(2\delta)$, integration of the Laurent expansion of the cotangent gives $I=\pi l-\delta-\pi\delta^2/6+O(\delta^3)$ and $J=-\pi\delta(l-1)+\delta^2/2+\pi\delta^3/18+O(\delta^4)$. Hence the linear term of $G$ is $\pi\delta(l/3-1/6)$; its leading pole is $2\pi/(3\delta)$. The static curvature term at this order is
\begin{equation}\label{eq:cuspbenchmark}
 \delta V_{\rm cyl}^{(2)}=\frac{\CF\CA\as^2 q^2r}{\pi}
 \left[\frac13\ln(2qr)-\frac16\right].
\end{equation}

From the EFT, the optical variables are $P=0,K=q^2$ and $\ell=r$. With $a_1/\CA=-8/3$, $\Tf n_f/\CA=11/4$, the soft coefficient from Eq.~\eqref{eq:staticmatch} is $-(\CF\CA\as^2q^2 r/4\pi)(8/9-4L_\nu/3)$. The exact free cylinder correlator along a static line is $G_E(t)=q^4/[16\pi^2\sinh^4(qt/2)]$, from the conformal image of the flat correlator. Its difference from the flat term has linear-detuning moment determined by
\begin{equation}
\begin{aligned}
 \FP_{s=2}\int_0^\infty\dd y\,\frac{y^{s-1}}{16\sinh^4(y/2)}&\\[-1ex]
 =\FP_{s=2}\frac{\Gamma(s)}6[\zeta(s-3)-\zeta(s-1)]&\\
 =-13/72.&
\end{aligned}
\end{equation}
The series $(16\sinh^4(y/2))^{-1}=\sum_{n\ge1}(n^3-n)e^{-ny}/6$ gives this Mellin transform in its convergence domain. Equivalently the finite part after the flat subtraction is $\lim_{\epsilon\to0}[\int_\epsilon^\infty(y/(16\sinh^4(y/2))-y^{-3})\dd y-\ln\epsilon/6]$. To fix the same dimensional scheme, differentiate the trace reference in Eq.~\eqref{eq:reference} with respect to $\Delta$, then add the convergent difference of the exact cylinder and its local flat-plus-$t^{-2}$ reference. In the limit $\Delta/q\to0$ the difference contributes $-\ln(\Delta/q)-\gamma_E-13/12$ in units $g^2\CF r^2 q^2/(6\pi^2)$ to the derivative. The reference derivative gives $\ln(2\Delta/\nu)+1+1/4$. The sum is $\ln(2q/\nu)-\gamma_E+1/6$. Multiplying by $\Delta$ gives the order-$\as^2 q^2 r$ ultrasoft term. Combining it with the soft coefficient cancels $\nu$ and reproduces Eq.~\eqref{eq:cuspbenchmark}. Terms of order $\as q^3r^2$ or higher in the small-distance expansion are not part of this comparison. The conformal isotropic result is thus already implicit in the cusp; the check constrains one trace combination of Table~\ref{tab:static}, not the trace-free or nonconformal matching.

\paragraph{Finite physical reference.} For the physical continuation of Eqs.~\eqref{eq:physicalDcurv} and \eqref{eq:physicalDacc}, write the coefficient of $c_b t^{-2b}$ as $-2\U_{ij}/3+\epsilon f_{1,ij}$, where
\begin{equation}
 f_{1,ij}=\tfrac53\T_{ij}+R_{ij}-\left(\tfrac R6+\tfrac23 a^2\right)\delta_{ij}-a_i a_j .
\end{equation}
At a time-independent central line, put $A=h_c-E-\ii0$ in each term. The subtraction is defined on the continued operator $A_d$ and the continued Coulomb gap in the common tree normalization before setting $d=3$ or removing EOM operators. If $A_d=A+\epsilon A_1+\cdots$, the $A_1$ contribution to a pole must be included when converting to a prescription which subtracts only the three-dimensional residue. For the cubic pole the linear term is $A^2A_1+AA_1A+A_1A^2$, without assuming commutativity. Source and potential conversions must use the same convention. In the continued-operator subtraction the remaining analytic reference is
\begin{align}
 \Sigma^{\rm ren}_{a,\U}(E)={}&\frac{g^2}{6\pi^2}\sum_c w_{ac}\U_{ij}
 r_i A\left[\ln\frac{2A}{\nu}-1\right]r_j\nonumber\\
 &+\frac{g^2}{8\pi^2}\sum_c w_{ac}f_{1,ij}r_i A r_j .\label{eq:physfinite}
\end{align}
Here the tensors are independent of the relative coordinate at the retained order. The second line is the evanescent tensor times the pole of $\Gamma(-1+2\epsilon)$. In the optical continuation replace $f_1$ by $2C_{i0j0}+(5K/3)\delta_{ij}$ in static optical variables; this recovers $c_P,c_K$ in Eq.~\eqref{eq:reference}. Equation~\eqref{eq:physfinite} is a finite local-reference kernel, to be supplemented by the actual state-dependent remainder, not a finite hard matching condition.
\subsection{Nonstationary propagation}\label{sec:real}
At leading light-field order we take a Hadamard vacuum or a color-invariant light preparation for which the transported connected covariance is color diagonal and the one-point field vanishes. The unreduced contour expression below also applies with a general matrix-valued connected covariance; the scalar weights $w_{ac}$ are appropriate only to the color-diagonal reduction. A nonzero mean field gives a separate term linear in the current in the influence action, and is not included in the two-point covariance. At two electric dipoles the connected influence on the closed time contour is
\begin{equation}\label{eq:SK}
 \log\mathcal F^{(2)}=-\frac{g^2}{2}\sum_{\eta,\eta'=\pm}\eta\eta'\int\dd t\,\dd t'\,j_\eta^A(t)G^{\eta\eta'}_{AB}(t,t')j_{\eta'}^B(t').
\end{equation}
The expression denotes the second-order connected insertion with contour ordering, not an ordinary commuting-current exponential for matrix-valued operators. The composite index includes color and electric polarization. The currents have the two endpoint transports and the complete intermediate propagation. The four contour components are $G^{++}=\theta G^>+(1-\theta)G^<$, $G^{--}=(1-\theta)G^>+\theta G^<$, $G^{+-}=G^<$, $G^{-+}=G^>$, with $\theta=\theta(t-t')$. They give $G_R=-\ii\theta(G^>-G^<)$ and $N=(G^>+G^<)/2$. With $j_\Delta=j_+-j_-$ and $j_c=(j_++j_-)/2$, Eq.~\eqref{eq:SK} is $-\ii g^2j_\Delta G_Rj_c-g^2 j_\Delta N j_\Delta/2$ in its contour ordering. Neither a Markov nor a secular approximation is required at this order. In particular the retarded pair self-energy contains $G^>$ times the intermediate channel evolution,
\begin{equation}
 \Sigma_a^R(t,t')=-\ii g^2\theta(t-t')\sum_c w_{ac}r_i G^>_{ij}(t,t')U_c(t,t')r_j .
\end{equation}
It is not merely the retarded light-field commutator. At higher light-field order the contour Wilson network cannot generally be replaced by a Gaussian scalar bath.

For a normalized initially factorized vacuum light preparation at a common finite regulator, resolve light out states $\lambda$ in the physical total-color sector. With interaction-picture coupling $V_I$, $K_\lambda=-\ii\int\dd t\langle\lambda|V_I(t)|0\rangle$ and $Z=\int_{t>t'}\dd t\dd t'\langle0|V_I(t)V_I(t')|0\rangle$. Completeness gives $Z+Z^\dagger=\sum_\lambda K_\lambda^\dagger K_\lambda$. The perturbative map with $K_0=1-Z-\ii\int\delta h_I$ preserves the norm through the retained order. Positivity holds for the full physical regulated evolution, not necessarily for a truncated density at arbitrary times. A correlated dressed initial state must be retained in the full Hilbert space and does not define a universal reduced completely positive map.

Minimal subtraction of the flat pole gives the reference kernel $-g^2\sum_c w_{ac}r_i J_3(h_c-E-\ii0;\nu)r_i/(6\pi^2)$, where
\begin{equation}\label{eq:J}
 J_3(B;\nu)=B^3[\ln(2B/\nu)-5/6].
\end{equation}
The time-ordered extension is made at common subtraction scale before separating real and virtual pieces. For example, for a smooth test matrix element $f(s)$ at positive lag, let $f_j=f^{(j)}(0)/j!$ and $n=2,4$. The finite part of the analytic one-sided distribution $s^{-n+2\epsilon}$, split at $l>0$, is
\begin{align}
 \mathrm{FP}_n[f]={}&\int_0^l\frac{\dd s}{s^n}
 \left(f(s)-\sum_{j=0}^{n-1}f_js^j\right)
 +\int_l^\infty\frac{\dd s}{s^n}f(s)\nonumber\\
 &+\sum_{j=0}^{n-2}\frac{f_j l^{j-n+1}}{j-n+1}
 +f_{n-1}\ln l .\label{eq:lagfinite}
\end{align}
Logs here are in a common reference unit. The derivative with respect to $l$ vanishes identically. The residue is $f_{n-1}/2$; the dimensional normalization replaces $\ln l$ by $\ln(\nu l/2)+\gamma_E$, with the additional finite terms from the continued electric tensor (or the factor $b^2$ in the flat term). The retarded local reference uses the analytically continued phase $\ii^{-n+2\epsilon}$, fixed by the causal boundary value of the stationary Laplace symbol. No analytic continuation of the actual time-dependent state is required. The test matrix elements contain the ordered heavy propagators and transported midpoint tensors, rather than instantaneous detunings. After local subtraction the remaining finite kernel depends on the full state and time evolution. This prescription acts on regulated smooth sources; sharp endpoints and singular heavy-operator domains require the corresponding source and contact matching. The two terms in Eq.~\eqref{eq:pole} are applied on both branches with opposite action signs. A local EOM change is a congruence on the inverse kernel, with the preparation and detector transformed by the same map. At a stable pole, for high block $H_h$, coupling $W_h$, and $R_i=(H_h-E_i)^{-1}$, the normalized embedding is
\begin{equation}\label{eq:embedding}
 I_i=\begin{pmatrix}\one\\-R_i W_h\end{pmatrix}(1+W_h^\dagger R_i^\dagger R_i W_h)^{-1/2}.
\end{equation}
A probe between stable poles is pulled back by $I_f^\dagger\mathcal B I_i$. Above threshold use the full spectral measure and its retarded boundary values, not an isolated pole normalization for a scattering state. For time-dependent unitaries $U$, $h\mapsto UhU^\dagger+\ii\dot U U^\dagger$, with both endpoint operators conjugated. The trace defining a normalized probability is then invariant. Color-gauge transformations cancel at all transported endpoints; at free field order the exterior derivatives kill the propagator's double-gradient gauge term. Metric Ward identities also vary the paths and endpoint sources. In particular for a curve functional with $T_C^{\mu\nu}=2\delta F_C/(\sqrt{|g|}\delta g_{\mu\nu})$ and force $f_\nu=\delta F_C/\delta X^\nu$, covariance yields $\nabla_\mu T_C^\mu{}_{\nu}=-\int\dd\tau f_\nu\delta^{(4)}(x-X)/\sqrt{|g|}$. The Ward identity does not fix finite transverse stress coefficients. Arbitrary finite-time inclusive predictions require matching the endpoint sources and the finite dynamical operator coefficients appropriate to the observable. The present real-time result is the leading two-dipole influence functional and its local renormalization, not a computed arbitrary-state transition probability.
\section{Applications to quarkonium spectroscopy}\label{sec:spectra}
\subsection{Perturbative spectral regime}\label{app:spectral}
The two ultrastatic spectra use a unit lapse, covariantly constant spatial curvature and zero electric rest tide, with distinct equal heavy flavors when excluding annihilation. The expansion is at fixed principal quantum number, $r\sim a_0$, $p\sim a_0^{-1}$ and $v\sim\as$. The leading stationary geometric energy is of order $R/M$; we seek its correction of order $\as R/M$. To exclude a contribution at this order one must match or count it, not merely classify its tensor structure.

At the hard scale the spatial kinetic coefficient is fixed by the kinetic mass, local Lorentz invariance and the covariant derivative. A spin-independent point insertion of curvature at order $1/M$ can only be a scalar multiple of $R/M$ or $R_{uu}/M$; the second vanishes here and the first cancels in a level difference at covariantly constant curvature. A contraction with trace-free Ricci curvature needs extra momentum or spin indices. The spin connection is fixed by the covariant kinetic term; one spin insertion has zero trace in a spin centroid and zero expectation in the spin singlet. Independent hard terms with curvature and two additional spatial momenta cost $p^2/M^2$.

For the soft one-insertion operator the isometric projection in Sec.~\ref{sec:action} fixes the second-order kinetic symbol and leaves a unitary connection and a multiplication term. At zero external gauge field the spin-independent singlet connection can be chosen locally zero by time reversal of the regulated static singlet. The multiplication term is the orthogonal squared derivative of the static singlet embedding. With a common soft regulator, differentiating the static eigenvalue equation gives $\langle m|\nabla_{\eta i}s\rangle=-\langle m|\mathcal F_{\eta i}|s\rangle/(E_m-E_s)$, where $\mathcal F_{\eta i}$ is the transported Hermitian force on endpoint $\eta$. In the orthogonal soft subspace this gives a force covariance with weight $\int_0^\infty\dd t\,t e^{-t(E_m-E_s)}=(E_m-E_s)^{-2}$. In the unit-lapse Wilson representation the recoil term is obtained from
\begin{equation}\label{eq:forceverify}
 \frac1{2M}\sum_{\eta=\pm}h_\eta^{ij}\int_0^\infty\dd t\,t
 \left[G_{\eta,ij}^{s,c}(t)-G_{\eta,ij}^{\rm iso}(t)\right]_{\rm soft}.
\end{equation}
Here the force includes $g$, the connected part removes the static-state insertion, and the subtraction includes the same isolated-line and ultrasoft prescription. This is the curved-space version of the normalized Wilson force formula \cite{firstrecoil,secondrecoil}. The energy normalization and endpoint transport are important for applying it.

The exclusion at order $g^2/M$ can be seen before any curvature expansion. For a regulated Gaussian propagator $D$ and a linear static current $J$, the normalized source functional with force test current $j$ is
\begin{equation}
 \log\frac{Z[J+j]}{Z[J]}=jDJ+\tfrac12 jDj,
\end{equation}
up to the conventional factors of $\ii$ in the Euclidean source. The connected second derivative is $D$, independently of $J$; hence the connected covariance of two linear field strengths is $\mathcal D_i D\mathcal D_j^\dagger$, independently of the other line. At order $g^2$ the color contraction on a force line is the isolated-line Casimir. Corrections to the nonlinear force, color transport or the Gaussian measure cannot contribute before order $g^4$: odd free-field contractions vanish and two extra gauge powers are required for the first connected interline contribution. The two same-line covariances in Eq.~\eqref{eq:forceverify} thus cancel pointwise at order $g^2$, including their curvature terms and any logarithms of $r$. This rules out a separation-dependent $\as R_{nn} f(\ln\mu r)/M$ in this projection, not merely a constant $f$. In the matching subspace the soft gaps are of order $1/r$, so inverse ultrasoft detunings cannot promote a $g^4$ coefficient. The free ultrasoft contribution is instead kept explicitly below. The first interline scalar recoil is of order $\as^2/(Mr^2)$ in flat space and $\as^2 R/M$ in the local linear-curvature term. Its finite curvature coefficients are not needed at the stated order.

For clarity the remaining counting at fixed $n$ is as follows. An additional hard recoil insertion in an interline exchange supplies at least $p/M$ on each transverse current or $1/(Mr)^2$ for a Darwin or retardation term; the first $1/M^2$ exchange energies are $O(\as v^2/r)$ and their geometric parts $O(\as v R/M)$. Mixed flat relativistic perturbations and flat $1/M$ potentials are relative $v^2$ in the wave function, whereas the ordinary one-loop static potential is relative $\as$ and must be retained. The geometric one-loop static potential is $O(\as^2 Rr)=O(\as R/M)$; two loops cost another $\as$. The geometric linear electric kernel contributes $O(\as r^2 R E_B)=O(\as R/M)$. The flat cubic ultrasoft term is $O(\as r^2 E_B^3)$; perturbing its heavy Hamiltonian or vertices by $O(Ra_0^2)$ changes it by $O(\as v^2 R/M)$. There is no enhanced mass tide on either ultrastatic slice. Additional Pauli/magnetic or higher-multipole vertices cost at least one power of $v$; an independent loop correction to the electric dipole costs another $\as$. These exclusions rely on the soft--ultrasoft separation, not an expansion in inverse ultrasoft gaps inside soft matching.

\paragraph{Leading low-frequency sector.} The local soft counting does not bound an unexpanded zero-detuning Wilson recoil graph in the infrared. In particular on $H^3$ the geometric part of the free vector-potential density is proportional to $q^2/\omega$, and a zero-gap transverse inverse may be singular. To see which low-frequency terms are represented by the retained relative E1 kernel, work at a common finite regulator. Transport the endpoint fields by the straight spokes. At leading multipole order the singlet has no monopole coupling; the relative one-field coupling is $g r^iE_i$ with its tree color matrix. In a velocity representation the kinetic connection and the rotation of the Coulomb potential supply the commutator with the full leading heavy Hamiltonian, not just its kinetic part. The quadratic term generated by this change of representation must be retained with the linear vertex.

For a single light mode of frequency $\omega>0$ and a heavy transition with gap $\beta>0$, choose a Hermitian generator $S$ such that the creation matrix element of the length interaction is $W_{L,mi}=\ii\omega S_{mi}$. Under $e^{\ii S}$ the velocity interaction is $W_V=W_L+\ii[S,H_0]$, hence $W_{V,mi}=-\ii\beta S_{mi}$. The second-order contact $\ii[S,W_L]-[S,[S,H_0]]/2$ has diagonal element $(\beta-\omega)|S_{mi}|^2$ per mode. The sum of the virtual and contact contributions is therefore
\begin{equation}\label{eq:iridentity}
 -\frac{\beta^2}{\omega+\beta}+\beta-\omega
 =-\frac{\omega^2}{\omega+\beta}.
\end{equation}
The common factor is $|S_{mi}|^2$. This identity accounts for the leading relative connection, potential rotation and associated contact in the one-mode low-energy sector without adding an independent zero-gap recoil term. In the hyperbolic vacuum the geometric integrand on the right is proportional to $q^2\omega/(\omega+\beta)$; it is integrable at zero. In the leading singlet Coulomb problem $\beta\ge E_B/n^2$, so the contribution of $0<\omega<\lambda$ is bounded by a constant times $\as a_0^2 q^2\lambda^2/(2E_B/n^2)$, uniformly as the frequency cutoff is removed. This bound follows from the positive octet spectral measure and finite dipole norm. In particular modes of frequency $O(q)$ in this sector give a correction suppressed by $O(q^2/E_B^2)$ relative to $\as q^2/M$ in the hierarchy of Sec.~\ref{sec:spectra}. The two-E1 expression retains the full $\omega+\beta$ denominator; expanding the gap at $\omega=0$ is not justified.

Equation~\eqref{eq:iridentity} is a regulated leading-dipole representation check, not a statement of unitary equivalence of unregulated vector-potential Fock spaces or an independent microscopic enumeration of every higher-order contact. Local independent two-field operators from the soft scale carry at least one additional multipole power relative to the iterated E1 coupling: for a field-strength-squared coefficient of size $g^2r^3$, its low-frequency geometric contribution is of order $\as r^3 R E_B^2$, smaller than $\as R/M$ by $E_Br$. In this count the eliminated soft gaps are of order $1/r$ and the ultrasoft gap is retained in the E1 iteration; singular higher-order low-energy graphs cannot be excluded by counting a zero-gap soft expression. The present spectral predictions are for the leading perturbative E1 sector with the stated matching; a full microscopic audit of subtractions and the uniformity of higher light-field orders is a separate test.

The enhanced common tide requires a separate power count. A flat Hamiltonian correction of size $E_Bv^j$ can give an energy response of order $v^{j-2}\T/M$ through perturbation by $F/E_B$. Thus a finite nonzero-tide spectrum at $O(\as\T/M)$ would require flat dynamics through relative $v^3$, in addition to matched geometric terms. No such spectrum is inferred from the leading local mobile pole. The zero-tide projections below avoid this enhancement.
Use a color-independent center trap with $\eta=a_0^2\langle P^2\rangle_0\ll\as$. The curvature derivative is taken first at fixed trap width and fixed nonzero coupling, before the small-$\eta$ limit. Choose a generic trap gap to avoid accidental coincidences with internal levels. Terms with a center derivative in the kinetic symbol are either odd in the center momentum or of relative order $\eta$; the leading isotropic $O(\eta R/M)$ term is retained explicitly. Take a spin centroid as the trace on the appropriate flat multiplet, or an isolated spin-singlet branch with curvature smaller than any relevant flat spin gap. For the $4P$ branch at $\beta_0>0$, the one-loop flat $4F-4P$ energy gap is
\begin{equation}
 E_{4F}-E_{4P}=-\frac{\as\beta_0 E_B}{\pi}\frac{13}{672}+O(\as^2 E_B).
\end{equation}
The stated curvature derivative is at fixed coupling, with the curvature insertion small relative to this splitting, rather than at the degenerate Coulomb point. The central flat perturbation does not mix $P$ and $F$. These restrictions close the operator counting for the stationary perturbative differences, not for generic finite mobile stress or time-dependent detector observables. In the linear-curvature coefficient the omitted local terms count as $O((\as^2+v^2)R/M)$ and omitted rest-projection terms as $O(\eta R/M)$, except for the explicit isotropic tree recoil correction below. At nonzero curvature, multiple geometric insertions are separately controlled by the gap of the chosen flat cluster. A smooth electrostatic remainder of order $\as q^3a_0^2$ is smaller than $\as q^2/M$ by $q/E_B$; the curvature derivative and the perturbative hierarchy must precede any extrapolation to finite curvature.
\subsection{Isotropic ultrastatic geometry}
We first consider $\mathbb R_t\times H^3_{1/q}$ in the static perturbative vacuum, with $q,\Lambda_{\rm QCD}\ll\EB$. The electric rest tide is zero and $\U=-q^2\one$. The free coexact spectrum \cite{camporesi} has the electric density $\delta_{ij}\omega(\omega^2+q^2)/(6\pi^2)$; this agrees with the local coefficient in Eq.~\eqref{eq:GE}. We consider the curvature derivative of the spin and magnetic centroid difference $E_{2S}-E_{2P}$ in a spherical color-independent center trap, subtracting the same flat trap. Take the derivative at fixed trap width before the rest limit $\eta=a_0^2\langle P^2\rangle_0\ll\as$. This specifies the center protocol and removes state-independent geometric recoil terms from the difference. For example, $U_L(\rho)=\rho^2/(4ML^4)$ in terms of center geodesic radius $\rho$ gives $\eta=3a_0^2/(2L^2)$ in the flat trap.

For $H^3_{1/q}$, the exact horizontal and vertical Jacobi factors transverse to the relative radius are $\cosh(qr/2)$ and $\sinh(qr/2)/(qr/2)$. Their product yields the Jacobian $[\sinh(qr)/(qr)]^2$. After conjugation to flat radial measure the relative kinetic plus Coulomb operator is
\begin{equation}
 -\frac1M\partial_r^2+\frac{q^2}{M}+\frac{q^2\ell(\ell+1)}{4M\sinh^2(qr/2)}-C_F\alpha_s q[\coth(qr)-1].
\end{equation}
The Coulomb term follows from the decaying radial Green function: the flux through a sphere of area $4\pi\sinh^2(qr)/q^2$ is constant. Its local expansion contains a smooth $O(q)$ constant assigned to the infrared reference. In the neutral channel it is common, since the constant part of the Coulomb Green function multiplies the square of the total color charge, including the ultrasoft gluon in an octet--gluon intermediate state. It must not be inserted as an isolated octet detuning. Expanding the displayed operator gives $-q^2\ell(\ell+1)/(12M)-C_F\alpha_s q^2r/3$, plus the common scalar $q^2/M$. With $\langle r\rangle_{2S}-\langle r\rangle_{2P}=a_0$, the relative difference is $-q^2/(2M)$. The center term $-q^2r^2[P^2-(P\cdot n)^2]/(16M)$ has spherical-trap average $-q^2\langle r^2\rangle\langle P^2\rangle/(24M)$. The two radial moments are $42a_0^2$ and $30a_0^2$, producing $-q^2\eta/(2M)$. This is the tree part of Eq.~\eqref{eq:H3}.

Put $k=C_F\alpha_s$ in this subsection. Use the definitions of \(\beta_0=(11\CA-4\Tf n_f)/3\), \(a_1=(31\CA-20\Tf n_f)/9\), \(L_\lambda(r)=\log(\lambda r)+\gamma_E\), and the optical continuation in Table~\ref{tab:static}. Substitution of the hyperbolic curvature values gives the curved interline singlet term
\begin{equation}\label{eq:soft}
\begin{aligned}
 \delta V_{s,q^2}=-k q^2r\biggl[\frac13+\frac{\as}{4\pi}\biggl\{&\frac{a_1+2\beta_0L_\mu}{3}\\
 &+\CA\left(-\frac{55}{9}+\frac43 L_\nu\right)\\
 &+\frac{20}{9}\Tf n_f\biggr\}\biggr].
\end{aligned}
\end{equation}
The flat one-loop perturbation is \(-k\as(a_1+2\beta_0L_\mu)/(4\pi r)\). In dimensionless radius \(x=r/a_0\), the singlet radial operator in units \(E_B\) is \(H_\ell=-\partial_x^2+\ell(\ell+1)/x^2-2/x\), with normalized radial state \(u_{n\ell}\) in \(\dd x\), explicitly
\[
u_{n\ell}(x)=\frac{2x}{n^2}\sqrt{\frac{(n-\ell-1)!}{(n+\ell)!}}\left(\frac{2x}{n}\right)^\ell e^{-x/n}L_{n-\ell-1}^{2\ell+1}(2x/n).
\] A constant multiplier of \(1/x\) simply rescales the Coulomb radius. Therefore the \(a_1+2\beta_0[\log(\mu a_0)+\gamma_E]\) response of the wave function cancels its direct factor in the linear curvature potential. The centrifugal and measure terms are diagonal in \(n,\ell\) in the \(q^2\) insertion, so at first order they have no radial wave-function correction.

For the nonconstant perturbation use the reduced radial solution
\[
 (H_\ell+1/n^2)y_{n\ell}=(x-\langle x\rangle_{n\ell})u_{n\ell},
 \qquad \langle u_{n\ell}|y_{n\ell}\rangle=0 .
\]
The first-order correction to \(\langle x\rangle\) per unit multiplier \(\epsilon\) of \(-2\log x/x\) is \(w_{n\ell}=4\langle y_{n\ell}|\log x/x|u_{n\ell}\rangle\). This sign follows from \((H_\ell-E_n)\delta u=-(\delta H-\delta E)u\). Write \(d_{n\ell}=\langle x\log x\rangle_{n\ell}\). The reduced-source polynomial solutions and their integrals are given in Appendix~\ref{app:matrix}. Put \(\Delta d=d_{20}-d_{21}=10/3-\gamma_E\), and \(\Delta(d+w)=11/2\). In units of \(\as q^2/(\pi M)\), the soft one-loop plus wave-function coefficient of the \(2S-2P\) double difference is therefore
\begin{equation}\label{eq:softexp}
 -\frac{11\beta_0}{6}
 -\frac{\CA}{2}\left[-\frac{55}{9}
                +\frac43\big(\log(\nu a_0)+10/3\big)\right]
 -\frac{10}{9}\Tf n_f .
\end{equation}
This calculation includes the mixed perturbation between the leading geometric Coulomb operator and the ordinary flat one-loop potential. A curvature coefficient formed using unperturbed Coulomb states only would depend on the arbitrary soft scale \(\mu\) at this order.

\subsubsection{Ultrasoft energy and spectral logarithm}\label{sec:bethe}
For hyperbolic \(H^3\), the free electric kernel is exactly \(G^E_{ij}=\delta_{ij}[\pi^{-2}t^{-4}+q^2/(6\pi^2t^2)]\) in the ultrastatic ground state. Its linear piece is pure optical trace and uses the finite constant \(c_K=1/4\). The intermediate Hamiltonian can be taken flat at this order within the geometric kernel. With \(A=h_o^{(0)}-E_n\), \(E_n=-E_B/n^2\), its renormalized contribution is
\[
 \delta E_{n\ell}^{\rm us}
  =-\frac{2\CF\as q^2}{3\pi}\,
     \left\langle r_i A\left[\log\frac{2(A-\ii0)}{\nu}+\frac14\right]r_i\right\rangle .
\]
At leading order \(A/E_B=\widehat A_n=-\nabla_x^2+2\gamma/x+1/n^2\), with \(\gamma=1/(N_c^2-1)>0\). Define \(\chi=\CA/\CF=2(1+\gamma)\). The dimensionless trace moment and logarithm are
\begin{align}
 m_{n\ell}&=\langle x_i\widehat A_n x_i\rangle
       =3+\chi\langle x\rangle_{n\ell},\nonumber\\
 b_{n\ell}&=\langle x_i\widehat A_n\log\widehat A_n\, x_i\rangle .\label{eq:b}
\end{align}
The unit \(a_0^2E_B=1/M\) makes \(m,b\) dimensionless. At \(n=2\) one has \(m_{20}-m_{21}=\chi\). The ultrasoft contribution in the same units as Eq.~\eqref{eq:softexp} is
\begin{equation}\label{eq:usdiff}
 -\frac{2\CA}{3}\left[\log\frac{2E_B}{\nu}+\frac14\right]
       -\frac{2\CF}{3}\Delta b,\qquad \Delta b=b_{20}-b_{21}.
\end{equation}
The kinetic \(3\) cancels the \(\nu\)-dependent state-independent term in this difference. It cannot be deleted from the functional calculus defining \(b\). In particular taking \(\widehat A_n=\chi/x\), the static detuning in units of \(E_B\), reproduces the gap part of the linear moment but gives a different logarithmic spectrum.

The radial logarithmic moment on a regular octet channel $L=\ell\pm1$ is
\begin{equation}
 b^L_{n\ell}=\langle f,\widehat A_{n,L}\ln\widehat A_{n,L}\,f\rangle,\qquad f=xu_{n\ell},
\end{equation}
where $\widehat A_{n,L}$ is the radial restriction of $\widehat A_n$. The sum defining $b_{n\ell}$ weights these moments by $\ell/(2\ell+1)$ and $(\ell+1)/(2\ell+1)$, omitting a channel of zero weight. Appendix~\ref{app:matrix} gives an explicit normalized Coulomb continuum representation. For convergence, at the origin $f=O(x^{\ell+2})$; applying $\widehat A_n$ gives $O(x^\ell)$, square-integrable for all the allowed channels (including $L=1$ from $\ell=0$). The regular boundary condition and exponential tail put $f$ in the operator domain of $\widehat A_n$. The spectral gap $\widehat A_n\ge n^{-2}>0$ and $|a\ln a|\le C_n(1+a^2)$ for $a\ge n^{-2}$ imply absolute convergence of each logarithmic moment. The spectral tail above $A$ obeys $\int_{a>A}a\ln a\,\dd\mu_f\le(\ln A)/A\,\|\widehat A_n f\|^2$ for $A>e$. Equivalently the same moment has an absolutely convergent subtracted resolvent representation
\begin{equation}
 b^L_{n\ell}=\int_0^\infty\dd t\left[\frac{\langle f,\widehat A_n f\rangle}{1+t}-\|f\|^2+t\langle f,(\widehat A_n+t)^{-1}f\rangle\right].
\end{equation}
The integrand is $\langle f,\widehat A_n[(1+t)^{-1}-(\widehat A_n+t)^{-1}]f\rangle$. Its large-$t$ absolute bound is $O(t^{-2})$ by the second spectral moment; the gap controls the lower endpoint. These arguments justify the cutoff removal and the spectral logarithms before evaluating them. In the isotropic difference the ultrasoft coefficient of $\ln\nu$ cancels Eq.~\eqref{eq:softexp}; since $2E_Ba_0=C_F\as$, the finite constants reduce to Eq.~\eqref{eq:H3}.

Combining the soft coefficient, the mixed flat one-loop wave-function perturbation and the finite ultrasoft matrix element gives, at fixed $M$,
\begin{align}
 \delta_q(E_{2S}-E_{2P})={}&-\frac{q^2(1+\eta)}{2M}\\
 &+\frac{\as q^2}{\pi M}\biggl[\frac{2\CA}{3}\left(\ln\frac1{\CF\as}+1\right)\nonumber\\
 &\quad-\frac{11\beta_0}{6}-\frac{10\Tf n_f}{9}\nonumber\\
 &\quad-\frac{2\CF}{3}\Delta b\biggr]+\cdots .\label{eq:H3}
\end{align}
The terms neglected include higher radiative orders and $O(\as\eta q^2/M)$. The factorization logarithm cancels between soft and ultrasoft terms; $\mu$ cancels through this order after including the flat perturbation. The result refers to the specified vacuum and trap. A time-dependent tidal field does not in general have this stationary spectrum. The logarithmic moments are specified by the convergent integrals above; no numerical spectrum is required for their definition.
\subsection{An anisotropic curvature combination}
For a covariantly constant axisymmetric spatial Ricci tensor on an ultrastatic slice, let $D_R=\operatorname{Ric}_{zz}-\operatorname{Ric}_{xx}$. Let $Q_n$ be the curvature part of the difference of the $|m|=1$ and $m=0$ spin-singlet $P$ energies, with the same trap and rest prescription. At leading order the relative Hamiltonian has tensor terms $-\operatorname{Ric}^{\rm STF}_{ij}L_iL_j/(12M)-\CF\as r\operatorname{Ric}^{\rm STF}_{nn}/12$. For the $P$ states the angular difference of $n_z^2$ is $-2/5$, and $x_n=(3n^2-2)/2$. Hence
\begin{equation}
 Q_n^{\rm tree}=\frac{D_R}{M}\left(-\frac1{12}+\frac{x_n}{15}\right)=\frac{D_R}{M}\frac{2n^2-3}{20}.
\end{equation}
The combination $\mathcal Q=7Q_2-12Q_3+5Q_4$ eliminates both leading tensor operators and finite static terms proportional to $r\operatorname{Ric}^{\rm STF}_{nn}$. In particular $\sum c_n=\sum c_n x_n=0$, $c=(7,-12,5)$. In the $4P$ sector take the derivative at the isolated flat QCD cluster, since the one-loop $4F-4P$ logarithmic moment difference is $13/672$ and removes the accidental Coulomb degeneracy for $\beta_0>0$. The curvature-induced energy shift must be smaller than this splitting. 
\subsubsection{Vacuum spectral curvature expansion on the product slice}
For $\mathbb R_t\times\mathbb R_z\times S^2_\kappa$ with $\kappa>0$, let $Y_{lm}$ be normalized sphere harmonics, $\lambda_l=\kappa l(l+1)$, $\omega^2=k^2+\lambda_l$. At $l\ge1$ the co-closed orthonormal spatial polarizations are
\begin{equation}
 e^{\ii kz}\frac{*_2\dd Y_{lm}}{\sqrt{\lambda_l}},\qquad
 e^{\ii kz}\left[\frac{\sqrt{\lambda_l}}{\omega}Y_{lm}\dd z+\frac{\ii k}{\omega\sqrt{\lambda_l}}\dd Y_{lm}\right].
\end{equation}
The $l=0$ traveling mode has no propagating transverse polarization. We choose the sector of zero background electric flux; no additional homogeneous electric field is populated. The scalar electrostatic $l=0$ mode is a constraint, not another transverse oscillator. For a neutral pair it can be fixed by zero flux at spatial infinity; an isolated globally colored line need not have finite energy on the cylinder. The isolated-line subtraction in the potential is a local common-regulator prescription, not a difference of such global colored energies. For a local scalar Green function the elliptic parametrix subtracts the full local singularity, leaving a smooth biscalar; at fixed midpoint its exchange-even expansion has a constant and then an $O(q^3r^2)$ term by dimensional scaling ($q=\sqrt\kappa$). In particular the $|z-z'|$ term of the separated scalar zero mode cannot be added to the local Coulomb parametrix by itself; the remaining scalar modes cancel its artificial short-distance cusp. The constant cancels for the neutral total color including an intermediate octet and its light field. Soft loop overlap terms are treated by Sec.~\ref{app:overlap}; the propagating low modes are retained in the following electric correlator. This argument is perturbative and does not describe nonperturbative flux sectors of the dimensionally reduced gauge theory. Let $w_l=\kappa(2l+1)/(4\pi)$ and $K(p)=\sum_{l\ge0}w_l e^{-p\lambda_l}$. The scalar zero-mode issue can also be checked directly along the axis. With $\rho,\rho_0>0$ axial separations, the neutral electrostatic Green-function difference is
\begin{equation}
 \begin{aligned}G(\rho)-G(\rho_0)&=\int_0^\infty\frac{\dd p}{\sqrt{4\pi p}}K(p)\\ &\quad\times\left(e^{-\rho^2/(4p)}-e^{-\rho_0^2/(4p)}\right).\end{aligned}
\end{equation}
The difference makes the constant large-$p$ heat mode integrable. Subtracting $K_0=(4\pi p)^{-1}$ and $K_1=\kappa/(12\pi)$ from $K$ gives, respectively, the flat Green difference and the linear term $-\kappa(\rho-\rho_0)/(24\pi)$. For the remainder use $|e^{-\rho^2/(4p)}-e^{-\rho_0^2/(4p)}|\le |\rho^2-\rho_0^2|/(4p)$. Since $K-K_0-K_1=O(\kappa^2p)$ for $\kappa p\le1$ and $O(\kappa)$ above it, the remainder is bounded by $C\kappa^{3/2}|\rho^2-\rho_0^2|$. Thus the axial expansion has precisely the coefficient $-\kappa r/(24\pi)$ of Eq.~\eqref{eq:frac} and no extra linear term from the global scalar mode.
The two electric correlators are
\begin{align}
 G_{zz}^E(t)&=\sum_{l\ge1}w_l\int\frac{\dd k}{2\pi}\frac{\lambda_l}{2\omega}e^{-t\omega},\\
 G_{AA}^E(t)&=\sum_{l\ge1}w_l\int\frac{\dd k}{2\pi}\frac{2k^2+\lambda_l}{4\omega}e^{-t\omega},
\end{align}
where $A$ is either orthonormal sphere direction. Subordination, $e^{-t\omega}/\omega=\pi^{-1/2}\int_0^\infty p^{-1/2}\exp[-p\omega^2-t^2/(4p)]\dd p$, and Gaussian $k$ integration give
\begin{align}
 G_{zz}^E(t)&=\frac1{4\pi}\int_0^\infty\frac{\dd p}{p}e^{-t^2/(4p)}[-K'(p)],\\
 G_{AA}^E(t)&=\frac1{8\pi}\int_0^\infty\frac{\dd p}{p}e^{-t^2/(4p)}[(K(p)-w_0)/p-K'(p)].
\end{align}
The sphere heat sum is $K(p)=(4\pi)^{-1}[p^{-1}+\kappa/3+O(\kappa^2p)]$, with derivative remainder $O(\kappa^2)$ for $0<\kappa p\le1$. This follows from the heat expansion on the unit sphere (or Euler--Maclaurin of $(2l+1)e^{-\kappa p l(l+1)}$), scaling and boundedness on closed positive heat-time intervals. In the radial correlator the constant term differentiates away. In the transverse one $K-w_0$ has constant $-2\kappa/(12\pi)$; the Gaussian integrals then give
\begin{align}
 G_{zz}^E(t)&=\frac1{\pi^2t^4}+R_z(t),\\
 G_{AA}^E(t)&=\frac1{\pi^2t^4}-\frac{\kappa}{12\pi^2t^2}+R_A(t).
\end{align}
For $\sqrt\kappa\,t\le1$, the small-$p$ remainder in either heat integral is bounded by $C\kappa^2\int_0^{1/\kappa}p^{-1}e^{-t^2/(4p)}\dd p$. For $\kappa p>1$, $K-w_0$ and $-K'$ decay exponentially in $\kappa p$; the subtracted local terms have integrable $p^{-3}$ and $\kappa p^{-2}$ tails. Thus $|R_i(t)|\le C'\kappa^2(1+|\ln(\kappa t^2)|)$ for $\sqrt\kappa t\le1$. For $\sqrt\kappa t\ge1$, scaling and the positive mode gap give $|R_i(t)|\le C''\kappa^2$. In the leading ultrasoft energy shift the flat octet resolvent has a positive gap $E_B/n^2$ and $\|r_i|n\ell\rangle\|^2=O(a_0^2)$ at fixed $n$. Integrating the remainder with this exponential bound gives $O(\alpha_s\kappa^2[1+|\ln(\kappa/E_B^2)|]/(M E_B^2))$. This is smaller than $\alpha_s\kappa/M$ in the stated curvature limit. It justifies using the linear geometric kernel for the product vacuum at the retained order without treating an arbitrary state as the vacuum. The leading flat-kernel variation with the curved heavy Hamiltonian is separately velocity suppressed by the counting above.

\subsubsection{Protected radiative combination}
The finite radiative value of \({\cal Q}\) can be computed from the same radial response and the product-slice vacuum. In the optical variables the traceless projection is \(\operatorname{Ric}^{\rm STF}_{nn}\), the Weyl restoration term is zero, and the extra soft factors in that projection are \(\CA(25/72-2L_\nu/3)+\Tf n_f(5/36)\). The fixed-projection conversion in Sec.~\ref{app:scheme} for an ultrastatic slice is \(e(n)=-R/9\), independent of direction. The direct and wave-function terms use the same $d_{n1},w_{n1}$ definitions as in the isotropic calculation. For \(n=3,4\) the reduced solution is generated by the same polynomial equation using \(\exp(-x/n)\) and a polynomial of degree \(n+2\). The corresponding exact integrals are
\begin{center}\small
\begin{tabular}{@{}lcc@{}}
\toprule
 & \(d_{n1}\) & \(d_{n1}+w_{n1}\)\\
\midrule
\(2P\) & \(137/12-5\gamma_E\) & \(27/4\)\\
\(3P\) & \(835/24+(25/2)[\log(3/2)-\gamma_E]\) & \(21\)\\
\(4P\) & \(1069/15+23[\log 2-\gamma_E]\) & \(513/12\)\\
\bottomrule
\end{tabular}
\end{center}
For example the reduced-source polynomial has coefficients given by solving \(L x^p=(2p/n-2)x^{p-1}+[\ell(\ell+1)-p(p-1)]x^{p-2}\), then removing its overlap with the input state. Therefore
\begin{align*}
 d_*&=\sum c_n d_{n1}\\ &=\frac{75}{4}+265\log 2-150\log 3,\\
 \sum c_n(d+w)_{n1}&=9.
\end{align*}
The \(x_z\) dipole connecting \(P\) to \(S\) has the squared-angular-coefficient difference \(-1/3\) between the two magnetic states, and the \(P\) to \(D\) one is \(-1/15\). This follows from the elementary \(\cos\theta Y_{\ell m}\) decomposition. Let \(b_n^{\rm an}=-b^0_{n1}/3-b^2_{n1}/15\). The exact linear moment of the same angular difference is \(-2\chi x_n/5\), from the tensor double commutator in Sec.~\ref{sec:us}, since the kinetic tensor term is independent of \(m\). Consequently the \(\nu\) logarithm already cancels in each \(Q_n\). In the weighted combination even the residual common scale and \(c_P=1/2\) pieces vanish. The weighted spectral moment is $b_*=\sum c_n b_n^{\rm an}$. The curvature logarithm in the direct soft potential gives $-2C_A d_*/15$, the mixed wave-function term and direct $\beta_0$ logarithm give $\beta_0\sum c_n(d+w)_{n1}/30=3\beta_0/10$, and the remaining ultrasoft term gives $C_F b_*/3$, in units $\alpha_s D_R/(\pi M)$. The local fixed-projection evanescent change for an ultrastatic slice is isotropic and cancels in the magnetic difference. The common finite constants and $\ln\nu$ terms are proportional to $\sum c_n x_n=0$. The coupling-scale derivative begins at the next radiative order. The resulting protected spectrum is
\begin{equation}\label{eq:Q}
 \mathcal Q=\frac{\as D_R}{\pi M}\left[-\frac{2\CA}{15}d_*+\frac{3\beta_0}{10}+\frac{\CF}{3}b_*\right].
\end{equation}
This projection does not require the general finite mobile curvature functions: state-independent terms and the relevant static finite constants cancel. On $\mathbb R_z\times S^2_\kappa$, $D_R=-\kappa$. A nonzero electric rest tide would invalidate this simplification.

The weighted combination also isolates an exact light-flavor response. For asymptotically free theories at fixed $N_c,M,\as(\mu)$ and the same curvature and trap protocol, the leading Coulomb operator and hence $b_*$ do not depend on $n_f$. Thus, comparing two theories with one additional massless light flavor at the same matched soft coupling,
\begin{equation}\label{eq:flavor}
 \mathcal Q(n_f+1)-\mathcal Q(n_f)=-\frac{\as D_R}{5\pi M}+\cdots .
\end{equation}
Equation~\eqref{eq:flavor} compares matched spectra in distinct theories, with the octet spectral logarithm eliminated. The flavor dependence originates in the QCD beta function. A leading Coulomb model at the same mass and coupling gives zero, whereas the radiative correction fixes a nonzero difference at the stated linear-curvature and center-rest order.
\section{Discussion}
A gravitational interaction common to the singlet and octet can be eliminated from the static color gap, but not from the cubic spectral moment of a mobile pair. In the latter case the mass tide enters through commutators with the relative kinetic energy. The sign of the geometric anomalous dimension need not, however, correspond to the sign of a curvature-induced energy shift.

The light-flavor dependence of the ultrastatic spectrum is a radiative effect. On $\mathbb R_z\times S^2_\kappa$, for example, an additional massless light flavor increases the weighted splitting by $\alpha_s\kappa/(5\pi M)$ at fixed matched mass and coupling. This difference is independent of the finite static constants and the octet spectral logarithm at the order considered.

The ultraviolet residue is local even when the geometry varies on the ultrasoft time scale. The finite response generally depends on the evolution of the heavy pair and the state of the light fields. For geometries with a nonzero electric mass tide, the curvature dependence of the stationary spectrum at radiative order remains sensitive to the finite mobile matching.

\section{Conclusions}
We have obtained the one-loop static singlet and projected-octet potentials and the leading geometric ultrasoft counterterms of a nonrelativistic heavy pair on a classical spacetime. The curvature dependence of the mobile anomalous dimension arises both from the chromoelectric correlator and from the gravitational interaction of the heavy constituents. The corresponding relative-dipole kernel has been formulated without an ultrasoft adiabatic approximation.

The calculation of the ultrastatic spectrum through $O(\alpha_s R/M)$ yields a combination of magnetic-sublevel splittings in which the tree curvature contributions and the relevant finite static constants cancel. At fixed matched mass, coupling and geometry, the change in this combination upon the inclusion of an additional massless light flavor is determined by the QCD beta function.

\appendix
\section{Operator-basis reduction and independent matching identities}\label{app:basis}
For an adjoint Hermitian argument $Y$, in addition to $\mathcal F$ and $\mathcal P_-$ defined in the text, write $\mathcal A(Y)=\tr(\mathscr S^\dagger Y\mathscr O+\mathscr O^\dagger Y\mathscr S)$ and $\mathcal D(Y)=\tr(\mathscr O^\dagger\{Y,\mathscr O\})/2$. Here $\mathcal P$ without a subscript denotes the pair field, not a color bilinear.
In a rest frame the invariant tensors are $\delta_{ij},\epsilon_{ijk},n^i$. Time integration by parts and local field redefinitions restore canonical first-order pair kinetics, leaving scalar Hamiltonian bilinears. Schur's lemma gives the two diagonal color projectors. A single adjoint couples the singlet and octet, and admits $d^{abc}$ and $f^{abc}$ octet contractions for $N_c\ge3$.

Take charge conjugation to send $\mathcal P(r)\mapsto \mathcal P^T(-r)$ and $F\mapsto-F^T$. A vector color current transforms like $F$ in color. Transposition reverses matrix order, so $\mathcal A,\mathcal P_-,\mathcal D$ require odd dependence on $n$ and $\mathcal F$ even dependence. The diagonal geometry slots require even dependence on $n$. Time reversal conjugates color amplitudes, with the electric field and acceleration time-even and the spatial curvature $F_{ij}$ of the gauge connection and spatial vector current time-odd. Thus $\mathcal P_-$ takes a time-odd argument and the other three Hermitian insertions take time-even arguments. The symmetry acts on the entire family of metrics and contours, including time-dependent ones.

\subsection{Geometric scalars}
The electric block $R_{i0j0}$ and the spatial Riemann block admit traces and projections along $n$. The spatial Riemann tensor in three dimensions is determined by its spatial-Ricci contractions. The mixed-time block gives no additional parity- and time-even scalar. The result can be written as $R,R_{uu},R_{nn},C_{nunu}$. The Ricci trace, timelike Ricci, longitudinal Ricci, and longitudinal electric Weyl components vary independently in local metric jets.
A single acceleration gives the exchange-odd $a_n$. Two accelerations give $a^2,a_n^2$. A single time derivative of acceleration gives no exchange-even scalar. The geometry jets at weight three are $\nabla R$, $aR$, $a^3$, $a\,\nabla_u a$ and $\nabla_u^2a$. Each has odd combined parity-times-time-reversal sign with an even number of relative unit vectors. For example the five indices of $\nabla R$ contribute one minus sign under the product of both reversals. The rest-frame acceleration and its first and second time derivatives have combined signs $-,+,-$. The nonrotating contour adds no angular-velocity spurion.

\subsection{Single fields and currents}
At weight two the parity-even contraction is $E_n$. At weight three the parity-even time-even gluonic contractions are the trace and longitudinal projection of $D_iE_j$ and the corresponding two contractions of $a_i E_j$. The parity-even time-odd gluonic contractions are $n^iD^jF_{ij}$, $n^ia^jF_{ij}$ and $n^i u^\alpha u^\beta e_i^\mu D_\alpha F_{\mu\beta}$. Gauge Bianchi does not tie the spatial magnetic curl to the time derivative of $E$.

For the full Fermi--Walker projection one has
\[
 D_\tau E_n=n^i u^\alpha u^\beta e_i^\mu D_\alpha F_{\mu\beta}
                    +n^i a^j F_{ij}.
\]
The derivative of $n^\mu$ is along $u^\mu$ and contracts to zero. With $T_X=\tr(\mathscr S^\dagger X\mathscr O)$, the leading pair equations give
\begin{align*}
 \partial_\tau T_X&=-\ii\Delta T_X+T_{D_\tau X},\\
 \mathcal P_-(D_\tau X)&\doteq-\Delta\mathcal A(X)\quad(X=g E_n).
\end{align*}
Here $\doteq$ denotes equality modulo the leading equations and an endpoint term. The derivative of the full dipole shifts its Wilson function and leaves the two magnetic contractions in the table. The local redefinition also transforms the two endpoint matching sources. The same canonical-kinetic convention removes a time derivative multiplying a diagonal Weyl term, with a corresponding shift of potentials and contacts. 

In the chiral limit the invariant flavor-summed vector and axial currents exhaust the dimension-three bilinears in color singlet or adjoint form. The axial contractions with $u,n$ are parity-odd. The color-singlet vector time component is charge-odd. Its longitudinal spatial contraction is time-odd even after inserting $n$ to make it charge-even. The adjoint vector components give $\mathcal F(J_u)$ and $\mathcal P_-(J_n)$. The bulk-field convention keeps the light-quark current jets independent of gluon derivative jets. Using the Yang--Mills equation would trade combinations of derivative and current operators, with contact shifts when a pair source is present.

Each of the two electric derivative projections and two electric acceleration projections can be varied independently. The spatial magnetic curl and the magnetic contraction with acceleration are independently variable as well. The dropped time derivative is the single integration-by-parts relation among the one-field scalar tensors at this order. Trying to remove a remaining imaginary vertex by division by the octet detuning generates further time derivatives of equal weight. The local expansion resolves the octet and retains its independent transition vertices.

\subsection{Canonical Coulomb-gauge check}\label{app:coulomb}
A separate evaluation uses the Gauss constraint on the ultrastatic optical slice, generalizing the flat canonical expansion \cite{coulombgauge}. Let $L=-\widetilde\nabla^2$, $G=L^{-1}$ and let $\mathfrak M=L+g\mathfrak M_1[A]$ denote the spatial Faddeev--Popov operator on transverse fields. The external-density quadratic term is $g^2(\rho,\mathfrak M^{-1}L\mathfrak M^{-1}\rho)/2$. Expanding the inverse yields $G-2gG\mathfrak M_1G+3g^2G\mathfrak M_1G\mathfrak M_1G$. The instantaneous term is the last kernel contracted in the transverse vacuum. The gluon and fermion charge-density screening terms follow from second-order perturbation of the cross term $g^2(\rho_{\rm ext},G\rho_{\rm light})$. Isolated-line pieces and vacuum normalization are subtracted. Both gluonic terms carry $C_AX_a$, and the fermion density carries $T_F n_f X_a$.

Let $L_1$ be the Hodge Laplacian and $H=e^{-t\sqrt L}/(2\sqrt L)$, $D=e^{-t\sqrt{L_1}}/(2\sqrt{L_1})$. The transverse kernel is $\mathsf T=D-d_xd_{y'}L^{-1}H$. Put $B_{ij'}=H_{;ij'}$, $Q(t)=\int_t^\infty B(s)\dd s$, $J=\int_0^\infty sB(s)\dd s$ and $B_G=G_{;ij'}=2Q(0)$. Then $\mathsf T(0)=D(0)-J$ and $\dot{\mathsf T}=\dot D+Q$. The instantaneous contraction is $3\mathsf T(0):B_G$; the screening kernel is $-2\int_0^\infty(\mathsf T:\ddot{\mathsf T}-\dot{\mathsf T}:\dot{\mathsf T})\dd t$. Since $\dot{\mathsf T}(0^+)=-P_T/2$ with $P_T=\mathbf1\delta_h-B_G$, integration by parts gives the combined kernel
\begin{equation}
 \mathscr C=4\mathsf T(0):B_G+4\int_0^\infty\dot{\mathsf T}:\dot{\mathsf T}\,\dd t-\tr\mathsf T(0;x,x)\delta_h(x,y).
\end{equation}
The correction is $X_a g^4 C_A[G\mathscr C G](x_+,x_-)$. The contact is retained for common-regulator comparisons, although scaleless in the homogeneous hard calculation. The color-density screening term cannot be omitted.

For an independent local metric insertion in spatial normal coordinates, the scalar variation is $\delta L=-R_{ikjl}z^kz^l\partial_i\partial_j/3+2S_{ij}z^j\partial_i/3$. Acting on a flat Gaussian heat kernel with one endpoint zero gives $-S_{ij}z^iz^j/(6u)$ times that kernel at heat time $u$. The Duhamel integral $-\int_0^s e^{-(s-u)L_0}\delta L e^{-uL_0}\dd u$ has Gaussian intermediate mean $ux/s$ and covariance $2u(s-u)\mathbf1/s$. It gives $K_L=K_0[1+S_{xx}/12+sS/6]$ in invariant measure. In radial parallel frames the connection term annihilates the radial Gaussian and the Hodge endomorphism adds $-s S_{ij}$ to the vector kernel. The Dirac square adds $-sS/4$ to the scalar coefficient for spinors. Mellin integration gives the local propagators of the text. This is an insertion of the quadratic metric jet at fixed proper separation, not an unrestricted off-shell stress vertex.

Set $\ell=1$ and remove $c_b$ from each propagator. For $f_p=(1+t^2)^{-p}$ and $J_p(t)=\int_t^\infty f_p(s)\dd s$, write $\dot D=\sum h_p t f_p$ and $B=\sum c_p f_p$. Use
\begin{align}
 J_p(0)&=\sqrt\pi\Gamma(p-1/2)/(2\Gamma(p)),\\
 \int_0^\infty t f_p J_q\dd t&=[J_q(0)-J_{p+q-1}(0)]/[2(p-1)],\\
 \int_0^\infty J_p J_q\dd t&=\int_0^\infty t f_p J_q\dd t+\int_0^\infty t f_q J_p\dd t,\\
 \int_0^\infty t^2 f_p f_q\dd t&=J_{p+q-1}(0)-J_{p+q}(0).
\end{align}
The terms proportional to $J_p(0)J_q(0)$ cancel the endpoint contribution. At separated points the remaining sums are
\begin{align}
 \mathscr C/c_b^2={}&4\sum h_p:h_q[J_{p+q-1}(0)-J_{p+q}(0)]\nonumber\\
 &-8\sum\frac{h_p:c_q}{2(p-1)}J_{p+q-1}(0)\nonumber\\
 &-4\sum c_p:c_q\left[\frac1{2(p-1)}+\frac1{2(q-1)}\right] \nonumber\\
 &\qquad\times J_{p+q-1}(0).
\end{align}
The arrays $h_p,c_p$ are obtained by differentiating $D,H$ as above; the radial Hessian is $-H_{\ell\ell}$ and the trace is the mixed Jacobi Hessian in Sec.~\ref{app:loops}. Contract an axial tensor with a general tensor via $A_{\perp}(\tr B-B_{nn})+A_{nn}B_{nn}$. Restoring $\ell$, write $\mathscr C=c_b^2\ell^{1-2d}(C_0+C_PP\ell^2+C_K K\ell^2)$. The sums give
\begin{align}
 C_0/j_d&=-8b^3(8b+3)/(2b+1),\\
 C_P/j_d&=-\frac{4b(8b^4-21b^3+5b^2-4)}{3(b-1)(2b+1)},\\
 C_K/j_d&=-\frac{8b^2(64b^4-48b^3-3b^2+15b-1)}{3(2b+1)(4b-1)}.
\end{align}
The two scalar legs attach by the same fractional Green identity as in Sec.~\ref{app:loops}. The normalized flat gluon bracket is $11/(3\epsilon)+31/9+22L_\ell/3$. The respective curvature ratios are
\begin{align}
 \rho_P&=\frac{8b^4-37b^3+13b^2-8}{12b(b-3)(b-2)(8b+3)}\nonumber\\
 &=-\frac1{11}+\frac{523}{2904}\epsilon+O(\epsilon^2),\\
 \rho_K&=\frac{64b^5-200b^4+172b^3+20b^2-53b+3}{3(b-2)(2b-3)(4b-5)(8b+3)}\nonumber\\
 &=-\frac2{11}+\frac{667}{363}\epsilon+O(\epsilon^2).
\end{align}
Multiplication gives precisely the gluonic pole and finite constants $25/72,55/9$ in Eq.~\eqref{eq:rawextra}. The fermionic density correlator is the same massless spinor contraction as in Sec.~\ref{app:loops}; it is independent of gauge fixing at this order. The canonical result tests the two local optical projections with the same physical contour and dimensional continuation, not a finite gravitational stress amplitude at arbitrary momentum.

\paragraph{Gauge and endpoint dependence.} A BRST gauge-parameter variation inserts a BRST-exact gauge-fixing operator. For a normalized closed singlet loop its expectation vanishes. For the covariant open color projection the variation acts at the endpoint indices, as $sU\propto c_fU-Uc_i$ with ghost $c$; the midpoint adjoint transporter carries the same endpoint variation. At fixed common BRST-compatible infrared regulator the hard connected endpoint insertions cluster at the endpoints and renormalize overlaps rather than the extensive hard pole. The same qualification applies to short spatial connector deformations. Isolated rest-line factors and cusp/endpoint terms are treated separately. This argument is for the normalized projected hard matching; it supplies no isolated observable octet energy.

\section{Auxiliary spectral matrix elements}\label{app:matrix}
\subsection{Reduced Coulomb matrix elements}
For \(n=2\), take \(u=e^{-x/2}p/\sqrt{\cal N}\), \(y=e^{-x/2}s/\sqrt{\cal N}\) with \((p,\mathcal{N})=(2x-x^2,8)\) or \((x^2,24)\). The exact solutions and integrals are
\begin{center}\small
\begin{tabular}{@{}lccc@{}}
\toprule
 & \(s(x)\) & \(d_{n\ell}\) & \(w_{n\ell}\)\\
\midrule
\(2S\) & \(-x^4/2+2x^3+12x^2-24x\) & \(59/4-6\gamma_E\) & \(-5/2+6\gamma_E\)\\
\(2P\) & \(x^4/2-15x^2\) & \(137/12-5\gamma_E\) & \(-14/3+5\gamma_E\)\\
\bottomrule
\end{tabular}
\end{center}
Indeed the polynomial operator maps \(x^p\) to \((p-2)x^{p-1}+[\ell(\ell+1)-p(p-1)]x^{p-2}\), and the integrals use \(\int_0^\infty x^m e^{-x}\log x\,\dd x=m!(H_m-\gamma_E)\). As a check, replacing \(\log x\) by \(1\) in the response gives \(-\langle x\rangle\) in both states.
\subsection{Continuum transform}
Use the dimensionless radial continuum state \(\sqrt{2/\pi}F_L(\gamma/k,kx)\) normalized with respect to \(\dd k\), where \(k\) in this subsection is the radial continuum momentum. \(F_L\) solves \(F_L''+(1-2\gamma/(k\rho)-L(L+1)/\rho^2)F_L=0\) as a function of \(\rho=kx\), and is regular at zero with unit sinusoidal amplitude at infinity. The repulsive Hamiltonian has no bound states. Standard Coulomb wave normalization gives \cite{dlmf}
\begin{align*}
 F_L(\eta_c,\rho)&=C_L(\eta_c)\rho^{L+1}e^{-\ii\rho}\\
 &\quad\times {}_1F_1(L+1-\ii\eta_c;2L+2;2\ii\rho),\\
 C_L(\eta_c)&=\frac{2^L e^{-\pi\eta_c/2}|\Gamma(L+1+\ii\eta_c)|}{\Gamma(2L+2)}.
\end{align*}
Here \(\eta_c=\gamma/k\), distinct from the center-trap parameter. The sign can be checked at the origin, where the subleading coefficient of \(F_L/\rho^{L+1}\) is \(+\eta_c/(L+1)\). Define
\begin{align*}
 I_{n\ell L}(k)&=\sqrt{\frac2\pi}\int_0^\infty x\,u_{n\ell}(x) F_L(\gamma/k,kx)\,\dd x,\\
 b^{L}_{n\ell}&=\int_0^\infty\dd k\,|I_{n\ell L}|^2\\
 &\quad\times(k^2+1/n^2)\log(k^2+1/n^2).
\end{align*}
Clebsch--Gordan reduction of \(n_iY_{\ell m}\) gives the summed angular weights \(\ell/(2\ell+1)\) and \((\ell+1)/(2\ell+1)\) for \(L=\ell-1,\ell+1\). Thus \(b_{20}=b^1_{20}\) and \(b_{21}=(b^0_{21}+2b^2_{21})/3\). The decomposition follows from the identity \(\sum_i n_i^2=1\) and the two adjacent-angular-momentum vector projections. It uses the same repulsive Coulomb continuum normalization as the flat QCD Bethe logarithm calculation of Ref.~\cite{kniehl}, applied to a linear detuning.

The radial transform is analytic before the final one-dimensional integral. For a monomial \(x^j\) in \(e^{x/n}u_{n\ell}(x)\), put \(p=j+L+2\), \(s=1/n+\ii k\), \(a=L+1-\ii\gamma/k\), \(c=2L+2\). For $u_{n\ell}=e^{-x/n}\sum_j p_j x^j$ with $\int |u_{n\ell}|^2\dd x=1$, the complete transform is $\sqrt{2/\pi}C_L(\gamma/k)k^{L+1}$ times the sum over $j$ of $p_j$ multiplying
\[
 \frac{\Gamma(p+1)}{s^{p+1}}
   {}_2F_1\left(a,p+1;c;\frac{2\ii k}{s}\right).
\]
Pfaff's transformation replaces the second parameter by \(-[p+1-c]\) and multiplies by \((1-2\ii k/s)^{-a}\). This small terminating polynomial representation fixes the continuation of the phase through \(k=0\). For repulsive color the low-\(k\) term is Gamow suppressed. The polynomial expression gives \(I=O(k^{-4})\) for \(2S\) and \(O(k^{-5})\) for an initial \(P\) state. These estimates agree with the spectral-domain convergence proof in Sec.~\ref{sec:bethe}.

\end{document}